\documentclass[a4paper,11pt]{article}
\pdfoutput=1

\usepackage{jheppub}

\usepackage{hyperref}
\usepackage{graphicx}
\usepackage{amsmath}
\usepackage{xcolor}
\usepackage{amssymb}
\usepackage[utf8]{inputenc}
\usepackage{subfigure}
\usepackage{bm}

\newcommand{\eq}[1]{eq.~\eqref{eq:#1}}
\newcommand{\eqs}[2]{eqs.~\eqref{eq:#1} and \eqref{eq:#2}}

\renewcommand{\sec}[1]{sec.~\ref{sec:#1}}

\newcommand{\subsec}[1]{sec.~\ref{subsec:#1}}

\newcommand{\fig}[1]{fig.~\ref{fig:#1}}

\newcommand{\app}[1]{app.~\ref{app:#1}}

\newcommand{\rcites}[1]{refs.~\cite{#1}}
\newcommand{\rcite}[1]{ref.~\cite{#1}}
\newcommand{\mathfig}[1]{\mathrm{fig.\ref{fig:#1}}}

\newcommand{\ord}[1]{{\mathcal O}(#1)}

\newcommand{\bra}[1]{\big\langle#1 \big\rvert}
\newcommand{\ket}[1]{\big\lvert#1 \big\rangle}

\newcommand{\bnslash}{\bar{n}\!\!\!\slash}
\newcommand{\bare}{\mathrm{bare}}
\newcommand{\bn}{\bar{n}}

\newcommand{\df}{\mathrm{d}}

\newcommand{\eps}{\epsilon}

\newcommand{\nn}{\nonumber}

\newcommand{\cD}{{\mathcal D}}
\newcommand{\cO}{{\mathcal O}}
\newcommand{\cL}{{\mathcal L}}

\newcommand{\gcusp}{\gamma_{\mathrm{cusp}}}

\newcommand{\Ecm}{E_\mathrm{cm}}

\graphicspath{ {figs/} }

\newcommand{\tr}{\mathrm{tr}}

\newcommand{\Tbar}{\overline{\mathrm{T}}}
\newcommand{\T}{\mathrm{T}}

\newcommand{\sumintX}{\sum \!\!\!\!\!\!\!\! \int\limits_{X}}

\newcommand{\ri}{\mathrm{i}}
\newcommand{\re}{\mathrm{e}}
\newcommand{\rd}{\mathrm{d}}
\newcommand{\rin}{\mathrm{in}}
\newcommand{\rout}{\mathrm{out}}

\DeclareMathAlphabet{\mathbbold}{U}{bbold}{m}{n}
\newcommand{\bbid}{\mathbbold{1}}

\allowdisplaybreaks[2]

\preprint{\begin{flushright}
MITP/20-058\\
FR-PHENO-2020-014
\end{flushright}}

\title{
Three-loop soft function for energetic electroweak boson production at hadron colliders
}

\author[a,b]{Ze Long Liu,}
\emailAdd{zelongliu@lanl.gov}
\author[c]{Maximilian Stahlhofen}
\emailAdd{maximilian.stahlhofen@physik.uni-freiburg.de}

\affiliation[a]{PRISMA$^+$ Cluster of Excellence \& Mainz Institute for Theoretical Physics\\
Johannes Gutenberg University, 55099 Mainz, Germany
}
\affiliation[b]{Theoretical Division, Los Alamos National Laboratory, Los Alamos, NM 87545, U.S.A.}

\affiliation[c]{Albert-Ludwigs-Universit\"at Freiburg, Physikalisches Institut, D-79104 Freiburg, Germany}

\abstract{
We calculate the three-loop soft function for the production of an electroweak boson (Higgs, $\gamma$, $W^\pm$, $Z$) with large transverse momentum at a hadron collider. 
It is the first time a soft function for a three-parton process is computed at next-to-next-to-next-to-leading order (N$^3$LO).
As a technical novelty, we perform the calculation in terms of forward-scattering-type loop diagrams rather than evaluating phase space integrals.
Our three-loop result contains color-tripole contributions and explicitly
confirms predictions on the universal infrared structure of QCD scattering amplitudes with three massless parton legs.
The soft function is a central ingredient in the factorized cross section for electroweak boson production near the kinematic endpoint (threshold), where the invariant mass of the recoiling hadronic radiation is small compared to its transverse momentum.
Our result is required for predictions of the near-threshold cross sections at N$^3$LO and for the resummation of threshold logarithms at primed next-to-next-to-next-to-leading logarithmic (N$^3$LL$^\prime$) accuracy.
}

\begin{document}
\maketitle

\section{Introduction}
The production of an electroweak (EW) boson (Higgs, $\gamma$, $W^\pm$ or $Z$) with sizable transverse momentum is among the most fundamental scattering processes at hadron colliders like the LHC. 
The corresponding cross sections are experimentally and theoretically relatively clean observables and allow precision phenomenology.
Their  measurements provide excellent tests of the Standard Model, and probe the parton distributions inside the colliding hadrons at small distances. 
EW gauge boson production at large transverse momenta also represents an important background to Higgs measurements and new physics searches.
On the other hand  the Higgs transverse momentum spectrum plays e.g.\ an important role for the analysis of the Higgs couplings.

\begin{figure}[t]
	\begin{center}
		\includegraphics[width=0.75\textwidth]{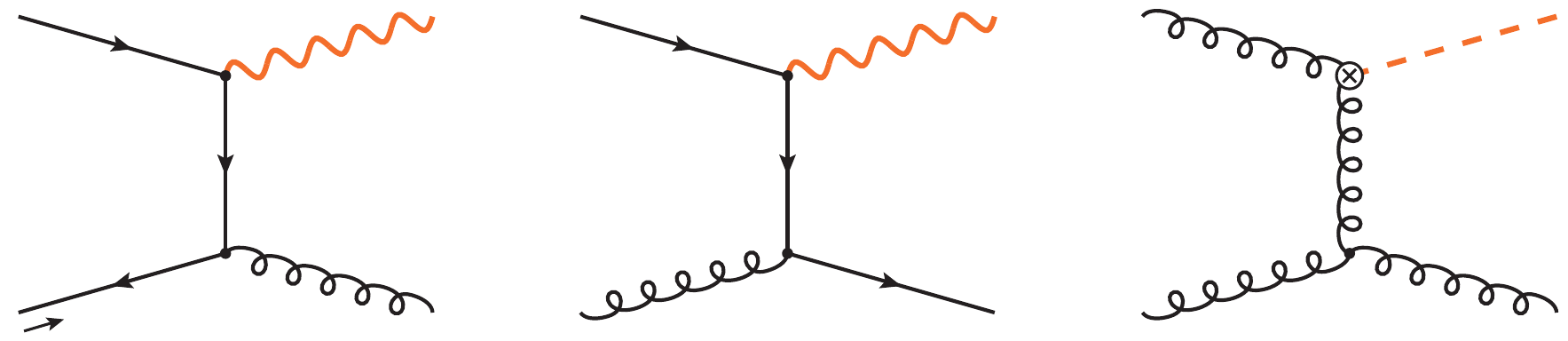}
	\end{center}
	\caption{Tree-level Feynman diagrams for electroweak boson production with large transverse momentum at hadron colliders. For EW gauge boson  (orange wiggly lines) production example diagrams of two different production channels are shown.
	\label{fig:treeVHj}}
\end{figure}

Moderate and large transverse momenta of EW bosons in hadron collisions are generated primarily by the recoil against hard QCD radiation (jets).
Corresponding tree-level processes with one hard final state parton are depicted in \fig{treeVHj}.
Compared to the inclusive EW boson production the leading order (LO) contribution therefore contains an additional power of the strong coupling constant $\alpha_s$, and it is technically harder to achieve the same level of accuracy on the theory side.
In the last couple of years  fixed-order QCD predictions for EW boson production with nonzero transverse momentum have reached NNLO precision~\cite{Boughezal:2013uia,Chen:2014gva,Boughezal:2015dva,Boughezal:2015dra,Boughezal:2015aha,Caola:2015wna,Ridder:2015dxa,Boughezal:2015ded,Ridder:2016nkl,Campbell:2016lzl,Chen:2016zka,Campbell:2017dqk,Gehrmann-DeRidder:2017mvr,Chen:2019zmr,Chen:2019wxf,Campbell:2019gmd}.
Depending on the kinematic region and the type of process the NNLO corrections can be substantial. For example the NNLO QCD corrections  to Higgs$+1$ jet production at the LHC enhance the cross section by roughly 20\%, while the uncertainties from scale variations still amount to about 10\%~\cite{Chen:2014gva}.
In view of future precision measurements a further reduction of the theory error would be desirable.

Whereas full QCD N$^3$LO results seem currently out of reach,
threshold approximations and the corresponding resummation of threshold logarithms can certainly improve the current state-of-the-art fixed-order predictions, at least at large transverse momenta.%
\footnote{Often the threshold terms amount to the bulk of the  corrections at a given order and the cross sections computed in the threshold limit represent indeed good approximations also at moderate transverse momenta.}
In the limit, where the transverse momentum of the EW boson becomes maximal (at fixed rapidity), the invariant mass of the recoiling hadronic radiation vanishes and vice versa.
Close to this threshold all final state QCD radiation is either soft or anti-collinear to the direction of the EW boson and the production cross section can be factorized into a convolution of hard, jet, and soft functions~\cite{Laenen:1998qw,Becher:2009th}.
Within the framework of soft-collinear effective theory (SCET)~\cite{Bauer:2000ew, Bauer:2000yr, 
Bauer:2001ct, Bauer:2001yt, Bauer:2002nz, Beneke:2002ph}
the jet and soft functions are expressed as effective operator matrix elements, while the hard function is a matching coefficient.
Each of the factorization functions obeys a corresponding renormalization group equation (RGE). Solving these and evolving the functions from the respective physical hard, jet, and soft scales to a common renormalization scale systematically resums large threshold logarithms (of the ratio between the transverse momentum and the hadronic invariant mass).
In this way near-threshold predictions for cross sections with NNLL and partial N$^3$LL resummation along with the corresponding NNLO threshold corrections were obtained~\cite{Becher:2009th,Becher:2011fc,Becher:2012xr,Becher:2013vva,Huang:2014mca,Becher:2014tsa,Schwartz:2016olw}.

In order to improve the predictions for EW boson production near threshold to N$^3$LO and (resummed) N$^3$LL$^\prime$ accuracy%
\footnote{For details and advantages of the primed counting, see e.g.\ \rcite{Almeida:2014uva}.}
the factorization functions are required at three-loop order.
Given the recent progress in the calculation of N$^3$LO four-particle scattering amplitudes in QCD~\cite{Jin:2019nya,Ahmed:2019qtg},
there is hope that also the three-loop virtual QCD corrections to EW boson + jet production, constituting the hard function, become available in the not-too-far future.
The relevant quark and gluon jet functions were computed at three loops in \rcites{Bruser:2018rad,Banerjee:2018ozf}.
The present paper is dedicated to the calculation of the three-loop soft function.

The soft function is given by a phase space integral constrained by the threshold measurement over a squared matrix element with one outgoing and two incoming Wilson lines along widely separated lightlike directions.
Although the relevant two-loop single~\cite{Dixon:2019lnw}, one-loop double~\cite{Zhu:2020ftr}, and tree-level triple emission~\cite{Catani:2019nqv} soft currents are available, performing the necessary phase space integrations is a difficult task.
We therefore follow a different approach.
Generalizing the `dispersive method' we used to calculate the two-parton soft function for heavy-to-light decays in \rcite{Bruser:2019yjk} we express our three-parton soft function in terms of a single Wilson line correlator corresponding to a forward scattering process. The phase space integrations then translate to loop integrations of Feynman diagrams with one internal and four external Wilson lines.
This trick allows us to directly apply modern multi-loop technology and makes the three-loop calculation of the soft function relatively straight-forward to carry out.

Our paper is organized as follows. 
In \sec{fac} we briefly discuss the factorization formula for the near-threshold production of EW bosons at the LHC and define the soft function employing the color space formalism.
In \sec{disc} we devise the `dispersive method'  to compute the soft function in terms of loop diagrams.
Section~\ref{sec:calc} describes the three-loop calculation in some detail and we present our results for the soft function in \sec{renorm}, including a discussion of its divergence and color structure. We conclude in \sec{conclu}.

\section{Threshold factorization and soft function definition}
\label{sec:fac}

In order to put the calculation performed in this paper into context we consider EW boson ($Y = \gamma, W, Z, h$) production at the LHC, i.e.\ the process $pp \to Y + X$.
The `hadronic' invariant mass of the final state particles in $X$ (including the proton remnants) is
\begin{equation}
  M_X^2 = (P_1 + P_2 -p_Y)^2\,,
\end{equation}
where $P_{1,2}$ are the proton momenta.
For $M_X \to 0$ the particles in $X$ form a collimated jet and any additional radiation must be soft. In this threshold limit the cross section factorizes, which allows the resummation of large logarithms $\sim \log(M_X/E_\mathrm{cm})$ with $E_\mathrm{cm}$ being the center of mass energy of the colliding protons~\cite{Laenen:1998qw}.

Defining the kinematic invariants at partonic level as
\begin{equation}\label{eq:stuparton}
\hat{s} = (x_1P_1+x_2P_2)^2=x_1x_2\Ecm^2\,,\quad
\hat{t} = (x_1P_1-p_Y)^2-m_Y^2\,,\quad
\hat{u}= (x_2P_2-p_Y)^2-m_Y^2\,,
\end{equation}
we can express the invariant hadronic mass as
\begin{align}
M_X^2= (\Ecm^2 + m_Y^2)\Bigl(1- \frac{m_\perp}{m_\perp^\mathrm{max}}\Bigr)
&=\left[(1-x_1)P_1 + (1-x_2) P_2 + p_X\right]^2\nn\\
&=m_X^2 + (\hat{s} + \hat{t})(1-x_1)+ (\hat{s} + \hat{u})(1-x_2) + \ord{M_X^4}\,,
\label{eq:MX}
\end{align}
where $m_X^2 =p_X^2$, with $p_X=x_1 P_1 + x_2 P_2 -p_Y$, is the total invariant mass of the final state partons (excluding the proton remnants),  $x_{1,2}$ are the fractions of the proton momenta carried by the incoming partons of the hard scattering process, $m_Y$ is the mass of the EW boson, and $m_\perp=\sqrt{p_T^2+m_Y^2}$ is its transverse mass.
The maximal transverse mass for fixed rapidity $y$ of the EW boson is given by 
$(m_\perp^\mathrm{max})^2 = (\Ecm^2+m_Y^2)^2/(2\Ecm \cosh y)^2$.
From \eq{MX} it is apparent that $M_X \to 0$ requires $m_X \to 0$ and $x_{1,2} \to 1$ simultaneously.
We also see that this limit is approached for large transverse momenta of $Y$.

The corresponding factorized near-threshold cross section as derived in SCET~\cite{Becher:2009th} for the channel $a b \to c Y$ with 
$\{a,b,c\} = \{q,\bar{q},g\}$, $\{q,g,q\}$,  $\{\bar{q},g,\bar{q}\}$, or $\{g,g,g\}$ takes the form%
\footnote{The factorization was performed here on the partonic level for small $m_X$. The factorized partonic cross section is then convoluted with the (threshold) PDFs to obtain \eq{xchadthres}.}
\begin{align}
\frac{\df^2 \sigma_{ab}}{\df p_T \df y} \propto{}&
\frac{\rd^2\sigma_{ab}}{\rd M_X^2 \rd y} \propto 
\int\!\! \rd x_1\int\!\! \rd x_2 \int\!\! \rd m^2 \int\!\! \rd \omega\, H_{ab} (\hat{s},\hat{t},\hat{u},m_Y,\mu)\,
f_{a}(x_1,\mu)f_{b}(x_2,\mu)  J_c(m^2,\mu) \nn\\
& \times  S_{ab}(n_{ij},\omega,\mu)\,
\delta\Bigl[M_X^2 - m^2 - 2E_J\omega 
- (\hat{s} + \hat{t} )(1-x_1) -(\hat{s} + \hat{u} )(1-x_2)\Bigr].
\label{eq:xchadthres}
\end{align}
The parton distribution functions (PDFs) $f_i(x,\mu)$ are evaluated close to their endpoints at $x=1$. The hard function $H_{ab}$ corresponds to a SCET Wilson coefficient obtained from matching SCET currents to full QCD matrix elements at the hard scale $\sim \Ecm \sim p_T \approx p_T^\mathrm{max}$ and contains the full QCD virtual corrections to the hard scattering process. 
It depends on the hard kinematics and is known to two-loop order for 
$Y =\gamma,W,Z,h$~\cite{Garland:2002ak,Gehrmann:2011ab,Jin:2018fak,Jin:2019ile}.
The jet function $J_c$ describes the collinear radiation along the jet direction $n_J^\mu$, which is anti-collinear to the direction of $Y$,%
\footnote{We define $n_i^2=0$ and $\bn_i\!\cdot\!n_i = 2$ for $i=1,2,J$.}
and only depends on the type of parton initiating the jet and its virtuality $\sim M_X$. The respective three-loop results were obtained in \rcites{Bruser:2018rad,Banerjee:2018ozf}.

The effects of soft wide-angle radiation (with momenta $\sim M_X^2/E_\mathrm{cm}$) are encoded in the soft function $S_{a b}$.
It depends on the color representation of the three hard partons and the total momentum component in $\bn_J^\mu$ direction of the soft partons. We also indicated a dependence on $n_{ij} \equiv n_i\!\cdot\! n_j$ with $i,j=1,2,J$, where the $n_i^\mu$ denotes the lightlike proton and jet directions in Minkowski space ($n_i^2=0$), respectively. This dependence is however such that it exactly cancels the dependence on the jet energy 
$E_J = [(\hat{s}+\hat{t})(\hat{s}+\hat{u})n_{12}/(2n_{1J}n_{2J}\hat{s})]^{1/2}$ 
in \eq{xchadthres} as can be made manifest by rescaling $\omega$~\cite{Becher:2009th}.
The soft function is the same for any $Y$ and was calculated at two loops in \rcite{Becher:2012za}.
In the present paper we compute it at three loops, i.e.\ $\ord{\alpha_s^3}$.

For later reference we also quote the Laplace transform of \eq{xchadthres}
\begin{align}\label{eq:xchadthreslap}
\frac{\rd^2\tilde \sigma_{ab}}{\rd Q^2\rd y}={}&
\int_0^\infty \!\!\rd M_X^2 
\exp\Bigl(-\frac{M_X^2}{Q^2e^{\gamma_E}}\Bigr)\frac{\rd^2\sigma_{ab}}{\rd M_X^2 \rd y}
\nn\\
\propto{}&
H_{ab}(\hat{s},\hat{t},\hat{u},m_Y,\mu)
\tilde f_a(\tau_1,\mu) \tilde f_b(\tau_2,\mu) \tilde j_c(Q^2,\mu) \tilde s_{ab}(\kappa,\mu)\,,
\end{align}
where the convolutions turned into simple products of $\tilde f_i$, $\tilde J_c$, and $\tilde s_{ab}$, which denote the PDFs, jet, and soft functions in Laplace space~\cite{Becher:2009th}, respectively, and
\begin{equation}\label{eq:varslaplace}
\tau_1 = \frac{Q^2}{ \hat{s} + \hat{t} }\,,\qquad
\tau_2 = \frac{Q^2}{ \hat{s} + \hat{u} }\,,\qquad
\kappa =\frac{Q^2}{2E_J}\,.
\end{equation}
In section~\ref{subsec:anomdim}, we derive the three-loop anomalous dimension of the soft function in Laplace space via the renormalization group (RG) invariance of \eq{xchadthreslap} and check the consistency with our explicit calculation.

In SCET the PDFs, the jet function, and the soft function are defined as operator matrix elements.
For illustration we quote here the relevant (unrenormalized) expressions for the quark and anti-quark PDFs of a proton ($p_{n_i}$) with momentum $P^-n_i^\mu/2$ employing the SCET label formalism~\cite{Bauer:2002nz,Stewart:2010qs}:
\begin{align}
f_q(x) ={}& \bra{p_{n_i}(P^-)}\theta(x) \,
\overline{\chi}_{n_i}(0) \frac{\bnslash_i}{2} 
\big[ \delta(x P^- - \overline{\mathcal{P}}_{n_i})\chi_{n_i}(0) \big]
\ket{p_{n_i}(P^-)}
\label{eq:PDF1}
\,, \\
f_{\bar{q}}(x) ={}& \bra{p_{n_i}(P^-)} \theta(x) \,
\mathrm{tr} \Big\{\frac{\bnslash_i}{2} \chi_{n_i}(0)  
\big[ \delta(xP^- - \overline{\mathcal{P}}_{n_i}) \overline{\chi}_{n_i}(0) \big] \Big\} 
\ket{p_{n_i}(P^-)}\,.
\label{eq:PDF2}
\end{align}
The composite quark field operator $\chi_{n_i}$ is gauge-invariant w.r.t.\ collinear gauge transformations and includes a collinear Wilson line in its definition, see e.g.\ \rcite{Stewart:2010qs} for details. The label momentum operator $\overline{\mathcal{P}}_{n_i}$ returns the total large light-cone (minus) momentum component of the $n_i$-collinear fields it acts on. A similar expression holds for the gluon PDF~\cite{Bauer:2002nz,Stewart:2010qs}.
The jet function and PDFs have trivial color structure, i.e.\ they do not have open color indices.

In contrast, the hard and soft functions can be regarded as color tensors with one index for each parton in the hard scattering amplitude and one for each parton in the complex conjugate amplitude. Corresponding color indices of soft and hard functions are pairwise contracted in the factorized cross section.
 (In \eqs{xchadthres}{xchadthreslap} color indices have been suppressed for the sake of compactness.)
A convenient way to deal with the colors structure of scattering processes with more than two partons is the color-space formalism~\cite{Catani:1996jh,Catani:1996vz}.
Applying it to our factorized cross section the soft function represents a square matrix in color space acting to the left and right on color vectors corresponding to the hard scattering amplitude and its complex conjugate.
In general we write for the hard function and the corresponding hard amplitude%
\footnote{The hard amplitude $C$ equals the QCD scattering amplitude with hard parton legs, where in the course of the SCET matching procedure, the IR divergences are subtracted.}
\begin{align}
  H_{a_1a_2\ldots b_1b_2\ldots} \propto 
  C_{a_1a_2\ldots}
  \Big(C_{b_1b_2\ldots}\Big)^*, \qquad
  C_{a_1a_2\ldots} = \sum_k \bra{a_1a_2\ldots}_k\, C_k,
\end{align}
where the $a_i$ (and $b_i$) are color indices of the parton legs and $\bra{a_1a_2\ldots}_k$ are (not necessarily orthonormal) vectors spanning the color space.
In the special case of three hard partons the color space is only one-dimensional
and we define%
\footnote{Here and in the following we often neglect the $\bar{q}g\to\bar{q}$ channel, because it is trivially related to the $qg\to q$ channel by charge conjugation. In QCD therefore all physical results for the two channels are the same.}
\begin{align}
  \bra{a_1a_2a_J} = 
  \ket{a_1a_2a_J}^* = 
  \begin{cases}
  t^{a_J}_{a_2a_1} & 
  \text{for the } q\bar{q} \to g \text{ channel,} \\
  t^{a_2}_{a_Ja_1} & 
  \text{for the } qg \to q \text{ channel,} \\
  \ri f^{a_2 a_J a_1} & 
  \text{for the } gg \to g \text{ channel,}
  \end{cases}
\end{align}
where $t^a_{bc}$ and $\ri f^{bac}$ are the $SU(N_c)$ generators in the fundamental and the adjoint representation, respectively.%
\footnote{The totally-symmetric tensor $d^{abc}=\tr[\{t^a,t^b\}t^c]/T_F$ is ruled out by charge conjugation symmetry of QCD as another possible color structure for the $gg\to g$ amplitude, see e.g.\ \rcite{Moult:2015aoa}.}
We thus have (for each production channel)
\begin{align}
  H_{a_1a_2a_Jb_1b_2b_J} S_{a_1a_2a_Jb_1b_2b_J}
  = H \, \bra{a_1a_2\ldots} \bm{S} \ket{a_1a_2a_J}
  = H S\, \big\langle a_1a_2a_J \big\rvert a_1a_2a_J \big\rangle\,,
  \label{eq:HScolor}
\end{align}
where $S_{a_1a_2\ldots}$ represents the soft function as a tensor with color indices and $\bm{S}$ represents the soft function as an operator acting on color space vectors, which is proportional to the unit operator ($\bm{S} = S \,\bbid$) in our one-dimensional color space.

Using this notation the soft function operator is defined by~\cite{Becher:2009th}
\begin{align}
  \bm{S}(\omega)=\big\langle 0 \big| \Tbar \big[ 
  Y_{J,\rout}^\dagger(0) Y_{1,\rin}^\dagger(0) Y_{2,\rin}^\dagger(0)
  \big]\,
  \delta(\omega-n_J\!\cdot \hat{p})\,
  \T \big[Y_{J,\rout}(0) Y_{1,\rin}(0) Y_{2,\rin}(0) \big]  
  \big| 0 \big\rangle\,,
  \label{eq:softfct1}
\end{align}
where $\T$ ($\Tbar$) indicates (anti-)time ordering, $\hat{p}^{\,\mu}$ is the soft momentum operator, and 
\begin{align}
  Y_{i,\rin} (x) &= \overline{\mathrm{P}} \exp \Bigl[- \ri g \! \int_{-\infty}^0 \!\!\df s\; 
  n_i\!\cdot\! A^c(x + sn_i)\, \bm{T}^c_i \Bigr]\,, 
  \label{eq:WLindef}\\
  Y_{i,\rout} (x) &= \mathrm{P} \exp \Bigl[+ \ri g \! \int_0^{\infty} \! \df s \; 
  n_i\!\cdot\! A^c(x + sn_i)\, \bm{T}^c_i \Bigr]\,,
  \label{eq:WLoutdef}
\end{align}
are the soft Wilson lines for an incoming and outgoing parton $i$, respectively.
The symbol $\mathrm{P}$ ($\overline{\mathrm{P}}$)denotes (anti-)path ordering of the color charge operators $\bm{T}^c_i$ and the associated (ultra)soft SCET gluon field operators $A^c_\mu(x)$.
The action of the $\bm{T}^c_i$ on color space vectors is defined by
\begin{align}
  \bra{a_1\ldots a_i \ldots a_n} \bm{T}^c_i \ket{a_1\ldots a_i\ldots a_n}
  \equiv
  \bra{a_1\ldots b_i \ldots a_n} T^c_{a_i b_i} \ket{a_1\ldots a_i\ldots a_n}
\end{align}
with 
\begin{align}
  T^c_{a_ib_i} \equiv
  \begin{cases}
    t^{c}_{a_ib_i} & 
    \text{if parton $i$ is a outgoing quark or incoming antiquark,}  \\
    -t^{c}_{b_ia_i} & 
    \text{if parton $i$ is an incoming quark or outgoing antiquark,}  \\
    \ri f^{a_i c b_i} & 
    \text{if parton $i$ is a gluon.}  \\
  \end{cases}
  \label{eq:ColorChargeOps}
\end{align}
Note that $\bm{T}^c_i \bm{T}^c_i = C_R \bbid$, where $C_R$ is the quadratic Casimir of the color representation $R$ of parton $i$ ($R=F$ for (anti)quarks and $R=A$ for gluons)
and 
\begin{align}
 \sum_{i=1}^n  \bra{a_1\ldots a_i\ldots a_n} \bm{T}^c_i = \sum_{i=1}^n \bm{T}^c_i \ket{a_1\ldots a_i\ldots a_n} = 0
\end{align}
due to color conservation in the scattering amplitudes.
Color charge operators of different partons commute trivially, i.e.\ $[\bm{T}_i^a,\bm{T}_j^b]=0$ for $i\neq j$, while $[\bm{T}_i^a,\bm{T}_i^b]=\ri f^{abc} \bm{T}_i^c$.

\section{Dispersive method}
\label{sec:disc}

The soft function defined in \eq{softfct1} represents an integral of a squared matrix element of one outgoing and two incoming  Wilson lines. This can be made explicit by inserting a complete set of (soft final) states ($X$),
\begin{align}
  \bm{S}(\omega)= \sumintX \,
  \delta(\omega-n_J\cdot p_X)\,
  \Big|\big\langle X \big| \T\big[Y_{J,\rout}(0) Y_{1,\rin}(0) Y_{2,\rin}(0) \big]  
  \big| 0 \big\rangle\Big|^2.
  \label{eq:softfct2}
\end{align}
Rather than evaluating phase space integrals we would like to 
compute the soft function in terms of forward-scattering-type loop diagrams.
A similar `dispersive' method was used for the calculation of the soft function for heavy-to-light decays near the kinematic endpoint as detailed in \rcite{Bruser:2019yjk}.

Following this approach, we will rewrite the $\delta$-function in \eq{softfct1} as a (momentum-space Wilson line) propagator-type expression using
\begin{align}
  \delta(x) = \mp \frac{1}{\pi} \,\mathrm{Im} 
  \biggl[ \frac{1}{x \pm \ri \delta}  \biggr]
  = \pm \frac{1}{\pi} \,\mathrm{Re} 
  \biggl[ \frac{\ri}{x \pm \ri \delta}  \biggr],
  \label{eq:deltaImRe}
\end{align}
where $\ri \delta$ represents an infinitesimally small positive imaginary part.
Our goal is to express the soft function in terms of a  single time-ordered (forward-scattering) matrix element, which can be evaluated using ordinary (Wilson line) momentum space Feynman rules.
As usual in momentum space perturbation theory the time-ordering is effectively implemented via the causal `$\ri0$' prescriptions for the propagators in the corresponding Feynman graphs.
Using \eq{deltaImRe} means that we will introduce another independent imaginary infinitesimal in the computation and special care has to be taken 
in order to avoid unphysical imaginary parts due to interference of $\ri\delta$ and $\ri 0$ in loop graphs giving rise to physical thresholds.%
\footnote{By `physical threshold' (not to be confused with the  large transverse momentum `threshold' of the EW boson) we refer to a branch cut along physical values of some kinematic invariant the soft function depends on besides $\omega$.
The corresponding amplitude for the heavy-to-light soft function discussed in \rcite{Bruser:2019yjk} is free of such thresholds.
The subtleties related to the difference of $\ri\delta$ and $\ri 0$ are therefore absent in that case and \eq{deltaImRe} can directly be applied with $\ri \delta \simeq \ri 0$.}

Concretely, we implement the $\delta$-function in \eq{softfct1} as follows
\begin{align}
  \bm{S}(\omega) =
  \big\langle O^\dagger \delta(\omega-n_J\!\cdot \hat{p})\, O \big\rangle &= \frac{1}{2\pi} \biggl(
  \big\langle O^\dagger \, \mathrm{Re} 
  \biggl[ \frac{\ri}{\omega-n_J\!\cdot \hat{p} + \ri \delta} \biggr] O \big\rangle
  -
  \big\langle O^\dagger \, \mathrm{Re} 
  \biggl[ \frac{\ri}{\omega-n_J\!\cdot \hat{p} - \ri \delta} \biggr] O \big\rangle
  \biggr) \nn \\
  &= \frac{1}{2\pi} \, \mathrm{Re} \biggl[
  \big\langle O^\dagger 
  \frac{\ri}{\omega-n_J\!\cdot \hat{p} + \ri \delta} \,O \big\rangle
  -
  \big\langle O^\dagger \frac{\ri}{\omega-n_J\!\cdot \hat{p} - \ri \delta} \, O \big\rangle
  \biggr], \nn\\
  &=  \frac{1}{2\pi} \, \mathrm{Re} \biggl[ \mathrm{Disc}_\omega
  \big\langle O^\dagger \frac{\ri}{\omega-n_J\!\cdot \hat{p}} \, O
  \big\rangle
  \biggr],
  \label{eq:ReRe}
\end{align}
where $O \equiv \T [Y_{J,\rout}(0) Y_{1,\rin}(0) Y_{2,\rin}(0)]$, 
$\langle \ldots \rangle \equiv \langle 0|\ldots|0\rangle$, and $\mathrm{Disc}_x\, g(x) \equiv \lim_{\beta \to 0} \,[ g(x + \ri \beta) - g(x - \ri \beta)]$.
At first sight taking the real part in \eq{ReRe} may seem redundant. It will however become important due to our treatment of $\ri \delta$ and $\ri 0$ in the following derivation.
We will illustrate the necessity of taking the real part in our dispersive approach for the soft function explicitly at two loops at the end of this section. 

Note that the operator matrix element in the last line of \eq{ReRe} alone is ambiguous. We therefore conveniently define
\begin{align}
  \sigma(\omega) \equiv \big\langle 0 \big| \Tbar \bigl[ 
  Y_{J,\rout}^\dagger (0) Y_{1,\rin}^\dagger (0) Y_{2,\rin}^\dagger (0)  \bigr]\,
  \frac{\ri}{\omega-n_J\!\cdot \hat{p} + \ri 0}\,
  \T \bigl[Y_{J,\rout}(0) Y_{1,\rin}(0) Y_{2,\rin}(0) \bigr]  
  \big| 0 \big\rangle.
  \label{eq:sigma}
\end{align}
Here we inserted $+\ri 0$ in the denominator in order to obtain a well-defined analytical expression for real $\omega$ and to make $\sigma(\omega)$ directly calculable via Feynman diagrams as we will demonstrate below.
With \eq{sigma} we can write
\begin{align}\label{eq:SdefDisc}
  \bm{S}(\omega)= S(\omega)\bbid = \frac{1}{2\pi} \, 
  \mathrm{Re} \Bigl[ \mathrm{Disc}_\omega \, \sigma(\omega) \Bigr] \,,
\end{align}
where the $\ri 0 \to 0$ limit inside $\sigma(\omega)$ is understood to be taken before the discontinuity.
Note that we could just as well define $\sigma(\omega)$ with $\ri 0 \to -\ri 0$, which would lead to $\mathrm{Disc}_\omega \, \sigma(\omega) \to [\mathrm{Disc}_\omega \, \sigma(\omega)]^*$.
Taking the real part in \eq{SdefDisc} removes this ambiguity and is therefore necessary to ensure the consistency with \eq{ReRe}.

Let us now consider the evaluation of $\sigma(\omega)$.
First, we notice that 
\begin{align}
  \sigma(\omega) = \big\langle 0 \big| \Tbar \bigl[ Y_{1,\rin}^\dagger(0) 
  Y_{2,\rin}^\dagger(0) \bigr] Y_{J,\rout}^\dagger(0) \,
  \frac{\ri}{\omega-n_J\!\cdot \hat{p} + \ri 0}\,
  Y_{J,\rout}(0) \,\T \bigl[ Y_{1,\rin}(0) Y_{2,\rin}(0) \bigr]  
  \big| 0 \big\rangle,
  \label{eq:sigma1}
\end{align}
because the gluon field operators $A_\mu^a$ inside $Y_{J,\rout}^{(\dagger)}$ are already (anti)time-ordered by default as a consequence of the (anti)path-ordering in \eq{WLoutdef}, and because the fields at positive times from $Y_{J,\rout}^{(\dagger)}$ are already to the left (right) of the fields at negative times in the $\T$ ($\Tbar$) product.
Using
\begin{align}
  \frac{\ri}{\omega-n_J\!\cdot \hat{p} + \ri 0} = 
  \int_{-\infty}^\infty  \df r \, 
  \re^{\ri (\omega - n_J\cdot \hat{p})r}\, \theta(r)
  =\int_0^\infty  \df r \, \re^{\ri (\omega - n_J\cdot \hat{p})r}\,,
\end{align}
we can thus write
\begin{align}
  &\sigma(\omega) = \int_0^\infty \!\df r\, \re^{\ri \omega r}
  \big\langle 0 \big| \Tbar \bigl[ Y_{1,\rin}^\dagger(0) 
  Y_{2,\rin}^\dagger(0)\bigr]  Y_{J,\rout}^\dagger(0) \,
  \re^{- \ri n_J\cdot\hat{p}\, r}\,
  Y_{J,\rout}(0) \,\T \bigl[ Y_{1,\rin}(0) Y_{2,\rin}(0) \bigr] 
  \big| 0 \big\rangle \nonumber\\
  &= \int_0^\infty \!\df r\, \re^{\ri \omega r}
  \big\langle 0 \big| \Tbar \bigl[ Y_{1,\rin}^\dagger(r n_J)
  Y_{2,\rin}^\dagger(r n_J)\bigr]  Y_{J,\rout}^\dagger(r n_J) \,
  Y_{J,\rout}(0) \,\T \bigl[ Y_{1,\rin}(0) Y_{2,\rin}(0) \bigr] 
  \big| 0 \big\rangle \nonumber\\
  &= \int_0^\infty \!\df r\, \re^{\ri \omega r}
  \big\langle 0 \big| \Tbar \bigl[ Y_{1,\rin}^\dagger(r n_J)
  Y_{2,\rin}^\dagger(r n_J) \bigr] 
  \mathrm{P} 
  \exp \Big[\ri g \!\int_{0}^{r} \!\! \df s \; n_J\!\cdot\! A^c(sn_J)\bm{T}^c_J \Big]
  \T \bigl[ Y_{1,\rin}(0) Y_{2,\rin}(0) \bigr] 
  \big| 0 \big\rangle \,.
  \label{eq:sigma2}
\end{align}
For an evaluation of $\sigma(\omega)$ via forward scattering Feynman diagrams all field operators in \eq{sigma2} must be time-ordered.
In the present form they are already time-ordered except for the operators in the $\Tbar$ product.
We can bring them into time order by the procedure sketched in the following, see also \rcite{Bruser:2019yjk} for a similar argument.

We first recombine $\sigma(\omega)$ with the $n_1$- and $n_2$-collinear matrix elements, i.e.\ the PDFs. We can now undo the BPS field redefinition~\cite{Bauer:2001yt} of the collinear fields, which made the decoupling of soft and collinear sectors at leading order in the SCET expansion manifest.
While the following arguments hold in general, let us consider the $q \bar{q}\to g$ channel for concreteness.
The resulting soft-collinear matrix element then has the schematic form
\begin{align}
  \big\langle p_{n_1} p_{n_2} \big| 
  \overline{\chi}_{n_1}(r n_J) \chi_{n_2}(r n_J)
  \mathrm{P} 
  \exp \Big[\ri g \!\int_{0}^{r} \!\! \df s \; n_J\!\cdot\! A^c(sn_J) \bm{T}^c_J \Big]
  \chi_{n_1}(0) \overline{\chi}_{n_2}(0) 
  \big| p_{n_1} p_{n_2} \big\rangle,
  \label{eq:softcollME}
\end{align}
where we suppressed the Dirac structure as well as the delta functions fixing the label momentum of the collinear ($\chi$) fields from \eqs{PDF1}{PDF2} for brevity.
As $r>0$ the fields in \eq{softcollME} are in fact time-ordered.
We can make this manifest by inserting the time-ordering operator $\T$ and redo the BPS field redefinitions $\chi_{n_1}\to Y_{1,\rin}\, \chi_{n_1}$ and $\overline{\chi}_{n_2}\to Y_{2,\rin}\, \overline{\chi}_{n_2}$.%
\footnote{To properly keep track of the color indices, see e.g.\ the detailed derivation of the factorization theorem based on the BPS field redefinition in \rcite{Becher:2009th}.}
Note, however, that the consistent interpretation of \eq{softcollME} as forward matrix element now requires that 
$\langle p_{n_1} p_{n_2} | = {}_\mathrm{out}\langle p_{n_1} p_{n_2} |$.
This entails that the field redefinition of the interpolating (anti)quark fields at time $+\infty$ generating part of the outgoing proton-proton state gives rise to an additional factor $Y_{1\infty} Y_{2\infty}$~\cite{Arnesen:2005nk} with
\begin{equation}
  Y_{i\infty}= 
    \overline{\mathrm{P}} \exp \Bigl[- \ri g \! \int_{-\infty}^{+\infty} \!\!\df s\; 
  n_i\!\cdot\! A^c(x + sn_i)\, \bm{T}^c_i \Bigr]
  =Y_{i,\rin}(rn_J) Y_{i,\rout}^\dagger(rn_J) 
\end{equation}
to the right of the $Y_{i,\rin}^\dagger$.
We emphasize that the color charge operator $\bm{T}_i$ in $Y_{i,\rout}$ is still that of an incoming parton $i=1,2$ w.r.t.\ \eq{ColorChargeOps}.

We can now factorize out the PDFs again, which are matrix elements of local operator products in SCET and as such unaffected by the time-ordering, and finally end up with
\begin{align}
  \sigma(\omega) = \!\int_0^\infty \!\! \df r\, \re^{\ri \omega r}
  \big\langle 0 \big|\T \Bigl[ Y_{1,\rout}^\dagger(r n_J)
  Y_{2,\rout}^\dagger(r n_J) 
  \mathrm{P}
  \exp \Big[\ri g \!\int_{0}^{r} \!\! \df s \, n_J\!\cdot\! A^c(sn_J) \bm{T}^c_J \Big]
  Y_{1,\rin}(0) Y_{2,\rin}(0) \Bigr]
  \big| 0 \big\rangle.
  \label{eq:sigmaT}
\end{align}
This expression for $\sigma(\omega)$ admits a straightforward evaluation in terms of forward-scattering-type loop diagrams using QCD Wilson-line Feynman rules in momentum space.

\begin{figure}[t]
  \begin{center}
    \includegraphics[width=0.25\textwidth]{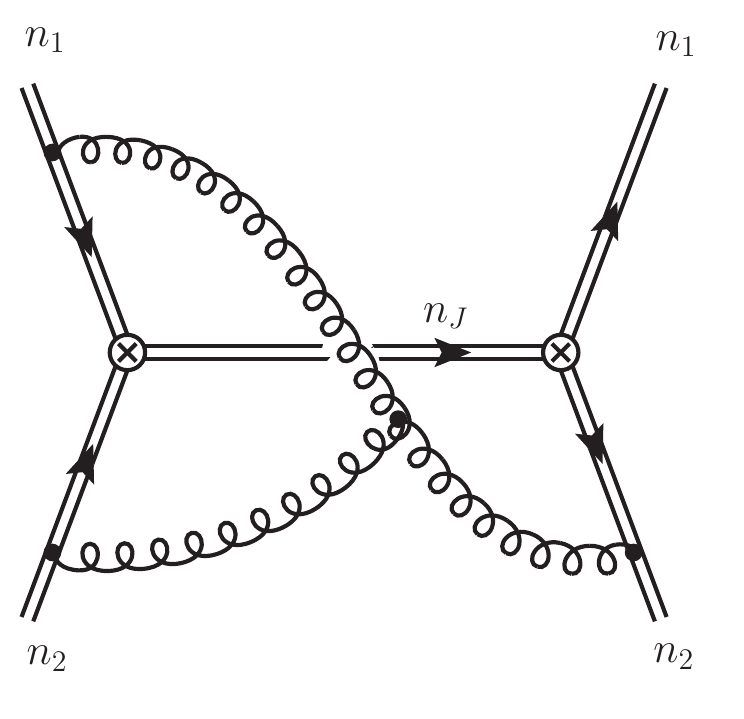}
  \end{center}
  \caption{Two-loop diagram contributing to $\sigma(\omega)$ and responsible for the imaginary part of $\mathrm{Disc}_\omega \, \sigma(\omega)$ at this order.
  The external double lines on the left (right) correspond to the semi-infinite lightlike Wilson lines $Y_{i,\rin(\rout)}^{(\dagger)}$, the middle double line corresponds to the finite-length Wilson line in \eq{sigmaT}.
  The little arrows indicate the parton flow for a generic parton channel with color charge operators $\bm{T}_i^c$ for each Wilson line vertex.
  The crossed vertices on the left (right) indicate the (complex conjugated) hard amplitude.
  \label{fig:2LD7}}
\end{figure}

One of the relevant Feynman graphs at two-loop order is shown in \fig{2LD7}.
Applying the Feynman rules and dimensional regularization ($d=4-2\epsilon$) we have
\begin{align}
 & \sigma(\omega)\big|_{\mathfig{2LD7}} = - 16 \pi ^2 \alpha_s^2 \, n_{12}\, f^{abc}\bm{T}^a_1 \bm{T}^c_2 \bm{T}^b_2  \int\!\! \frac{\df^dl_1}{(2\pi)^d}
  \frac{\df^dl_2}{(2\pi)^d} \nn\\
 &\times \frac{n_2\!\cdot\! (l_1+l_2)}{
  	[n_1\!\cdot\!(l_1\!-\!l_2)+\ri 0](n_2\!\cdot\!l_2+\ri 0)(n_J\!\cdot\! l_1+\omega+\ri 0) (n_2\!\cdot\!l_1+\ri 0)
  	(l_1^2+\ri 0)(l_2^2+\ri 0)[(l_1\!-\!l_2)^2+\ri 0]}
  \nn\\
& =-2\ri\, C_S C_A \bbid \left(\frac{\alpha_s}{4\pi}\right)^2 
(2 \pi)^{2\epsilon}\,(-\omega-\ri 0)^{-1-4\epsilon}
(n_{1J}n_{2J})^{2\epsilon} 
\Biggl[
(n_{12})^{-2\epsilon} \,\Gamma (1-2 \epsilon ) \Gamma^2 (-\epsilon ) \Gamma (2 \epsilon ) \Gamma (4 \epsilon ) 
\nn\\
&\quad
+ (-n_{12}-\ri 0)^{-2\epsilon} \frac{\Gamma^2 (-2 \epsilon ) \Gamma^2 (-\epsilon )
	\Gamma (2 \epsilon ) \Gamma (1+4 \epsilon ) \Gamma (1+3 \epsilon)}{\Gamma (\epsilon )}\Biggr].
\label{eq:D7calc}
\end{align}
The left-right and up-down mirror graphs yield the exact same result.
The color factor $C_S$ depends on the channel and is given in \eq{csdef}.
To obtain the corresponding contribution to the soft function according to \eq{SdefDisc} we first take the discontinuity in $\omega$ (after $\omega +\ri0 \to \omega$) using
\begin{equation}
  \mathrm{Disc}_\omega (-\omega)^{-1-a\eps} 
  = -2 \ri \sin(\pi a \eps) \omega^{-1-a\eps} \theta(\omega)\,.
  \label{eq:TakeDisc}
\end{equation}
The resulting expression has an imaginary part from the branch cut in the $(-n_{12}-\ri0)^{-2\eps}$ factor for (physical) $n_{12}>0$ (related to a physical threshold in the full QCD one-gluon emission amplitude for the process $ab\to c Y$).
In a calculation based on \eq{softfct2} as performed e.g.\ in \rcite{Becher:2012za} such imaginary parts also appear in the soft amplitude, but cancel in the product with the complex conjugated amplitude.
Equivalently, using two-loop diagrams with a unitarity cut (through a single gluon line) the imaginary part cancels between the diagram corresponding to \fig{2LD7} and its left-right mirror diagram.
In our approach the imaginary part is removed by explicitly by taking the real part in \eq{SdefDisc}.
At two loops the diagram in \fig{2LD7} (together with its mirror graphs) is the only one with an imaginary part after taking the discontinuity.
At three loops we found quite a number of such diagrams.

\section{Calculation}\label{sec:calc}

In this section, we present details of our three-loop calculation of the soft function based on \eqs{SdefDisc}{sigmaT}. The calculation is performed in general covariant gauge with gauge parameter $\xi$, where $\xi=0$ corresponds to Feynman gauge. Ultraviolet (UV) and infrared (IR) divergences are regulated using dimensional regularization ($d=4-2\epsilon$).

\begin{figure}[t]
	\begin{center}
		\subfigure[]{\label{fig:triplea}\includegraphics[width=0.23\textwidth]{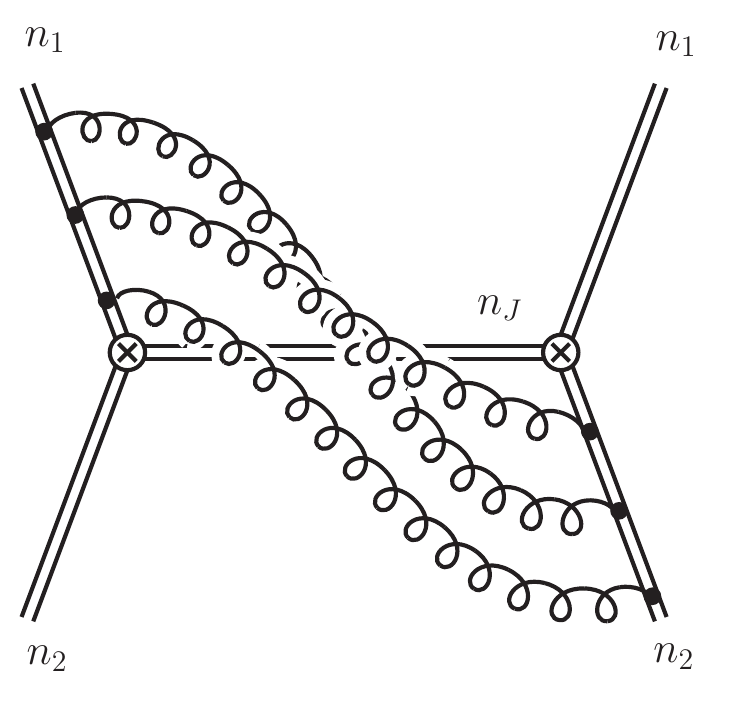}}\quad
		\subfigure[]{\label{fig:tripleb}\includegraphics[width=0.23\textwidth]{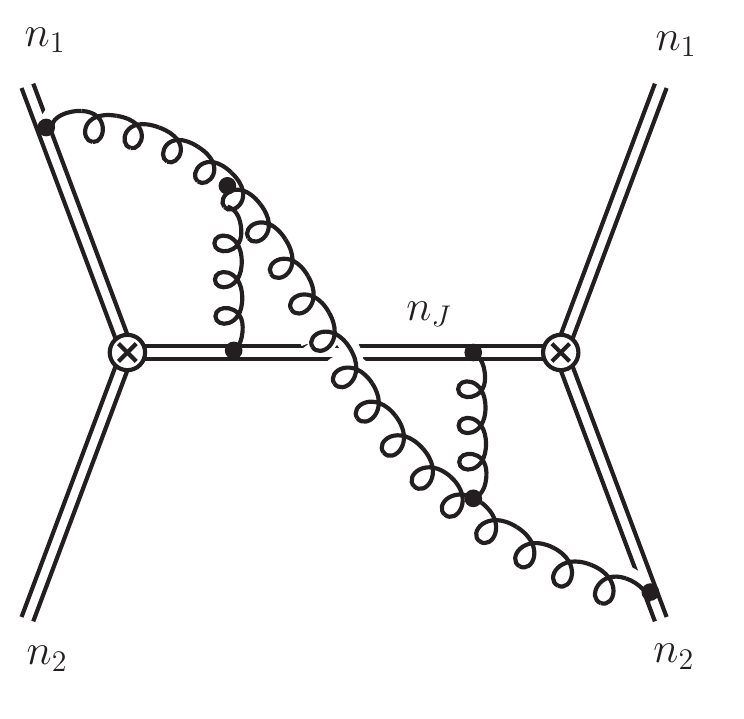}}\quad
		\subfigure[]{\label{fig:triplec}\includegraphics[width=0.23\textwidth]{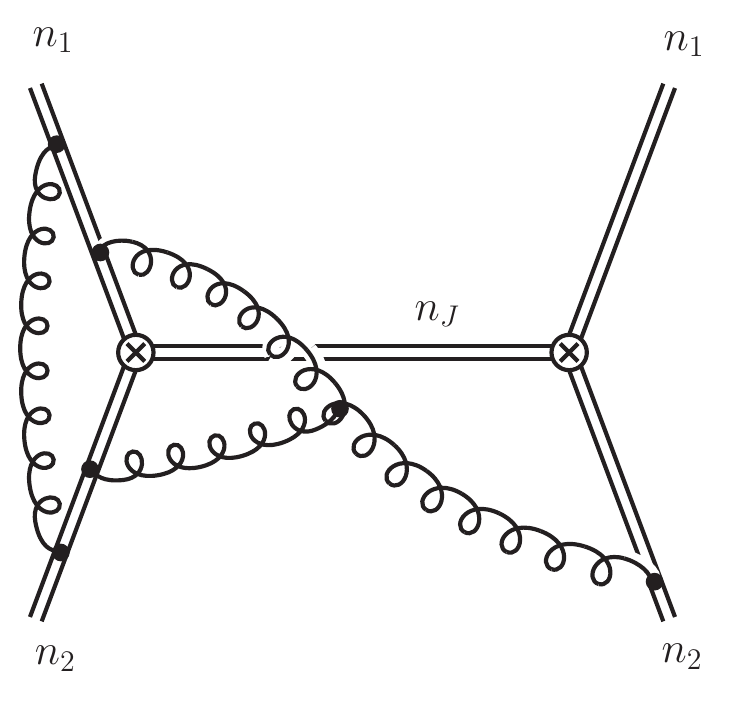}}\quad
		\subfigure[]{\label{fig:triplec}\includegraphics[width=0.23\textwidth]{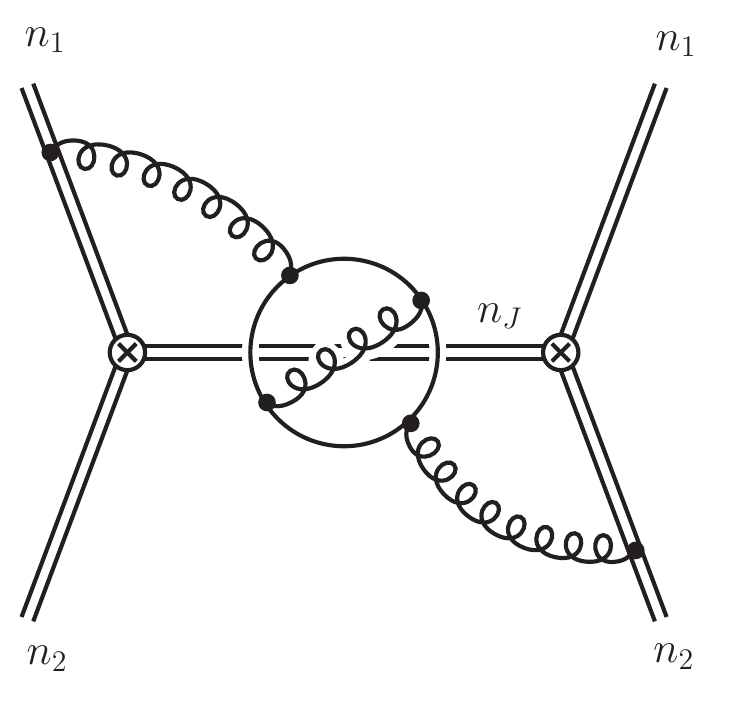}}
	\end{center}
	\caption{\label{fig:feyndia} Examples for three-loop Feynman diagrams that contribute to the soft function $S(\omega)$.}
\end{figure}

\begin{figure}[t]
  \begin{center}
    \includegraphics[width=0.23\textwidth]{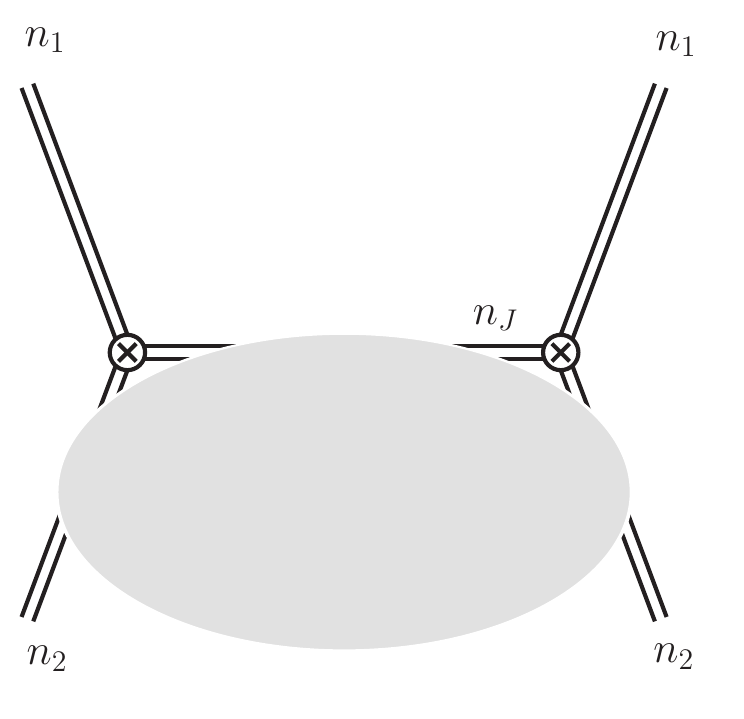}
    \qquad\qquad
    \includegraphics[width=0.23\textwidth]{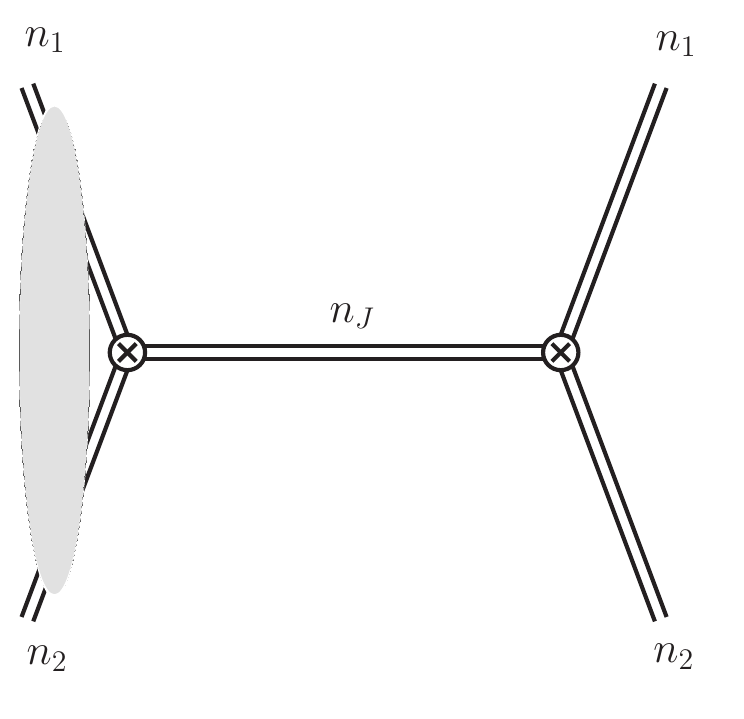}
    \qquad\qquad
    \includegraphics[width=0.23\textwidth]{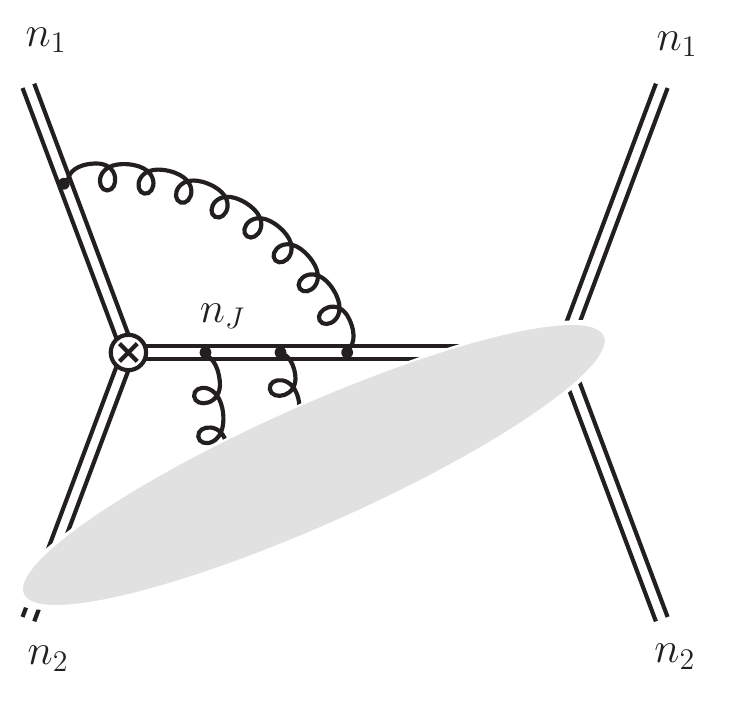}
  \end{center}
  \caption{\label{fig:feyndia0}
  Using the same arguments that lead to \eq{Gres} one can show that (sub)diagrams whose integrand is independent of at least one of the lightlike vectors $n_i$ and any external invariant $l_j^2$ (where $l_j$ is an external loop momentum w.r.t.\ the subloop) vanish in dimensional regularization.
  Examples for such diagrams are shown in this figure. The gray blob stands for all possible subdiagrams that involve the Wilson lines it overlaps.
}  
\end{figure}

We generate the relevant three-loop Feynman diagrams with \texttt{qgraf}~\cite{Nogueira:1991ex}. Some examples of non-trivial three-loop graphs are shown in fig.~\ref{fig:feyndia}. 
Based on dimensional counting and investigating the behavior of the corresponding loop integrals under rescaling
\begin{equation}
n_1\to \lambda_1 n_1\,,\qquad
n_2\to \lambda_2 n_2\,,\qquad
n_J\to \lambda_J n_J\,,\qquad
\omega\to \lambda_J \omega\,,
\label{eq:rescalings}
\end{equation}
we can directly identify already at this stage classes of diagrams that vanish as scaleless integrals, as illustrated in \fig{feyndia0}.
We are then left with 1290 non-trivial diagrams to be computed. 

These diagrams are further processed by a private $\texttt{Mathematica}$ code~\cite{RobinThesis} which first assigns the momentum space Feynman rules and performs simplifications of the Dirac, Lorentz and color structure. 
After that the diagrams are expressed as linear combinations of scalar Feynman integrals. 
Exploiting the $n_1\leftrightarrow n_2$ interchange symmetry these can be mapped onto twenty integral families corresponding to different sets of fifteen linearly independent linear (Wilson line) and quadratic propagators.
The mapping of scalar integrals to specific families requires partial-fraction decomposition of linear propagators followed by suitable shifts of the loop momenta. 
In order to automate the extensive partial fractioning we employed the algorithm described in \rcite{Pak:2011xt}.

Next, we use the public program $\texttt{FIRE5}$~\cite{Smirnov:2014hma} to perform the integration-by-parts (IBP) reduction of the integrals in each of the twenty integral families to a set of master integrals (MIs).
In three of the families we find an additional partial-fraction identity among the MIs returned by $\texttt{FIRE5}$, reducing their number by one, respectively.
We then identify redundant MIs across the families, i.e.\ subsets of MIs with the same integrand up to loop momentum shifts despite being members of different families.
Finally, the sum of all three-loop diagrams contributing to $\sigma(\omega)$ can be expressed as a linear combination (with real $d$-dependent coefficients) of 73 linearly independent MIs belonging to eleven different integral families. 
In this expression the gauge parameter $\xi$ manifestly cancels out, which represents a strong check of our setup. 

The integrals in each family can be written as
\begin{align}
G(\vec{a}, \vec{b},\vec{c}, \vec{e},\epsilon) = \bigl(\ri\pi^{\frac{d}{2}}\bigr)^{-3}
\!\int\!\!
\frac{
	\df^dk_1\,\df^dk_2\,\df^dk_3}{
	\cD_1^{a_1}
	\cD_2^{a_2}
	\cD_3^{a_3}
	\cD_4^{a_4}
	\cD_5^{a_5}
	\cD_6^{a_6}
	\cD_7^{b_1}
	\cD_8^{b_2}
	\cD_9^{b_3}
	\cD_{10}^{c_1}
	\cD_{11}^{c_2}
	\cD_{12}^{c_3}
	\cD_{13}^{e_1}
	\cD_{14}^{e_2}
	\cD_{15}^{e_3}}\,,
\label{eq:Gintegral}
\end{align}
where the denominators $\cD_{1,\cdots ,6}$ correspond to quadratic propagators, and $\cD_{7,8,9}$, $\cD_{10,11,12}$, and $\cD_{13,14,15}$ correspond to propagators of Wilson lines along the $n_1$, $n_2$ and $n_J$ directions, respectively. 
To give an example, one of our integral families is defined by
\begin{align}
&\cD_1=-k_1^2\,, 
& &\cD_2=-k_2^2\,,
& &\cD_3=-k_3^2\,, & \nn\\
& \cD_4=-(k_1-k_2)^2\,,
& &\cD_5=-(k_2-k_3)^2\,, 
& &\cD_6=-(k_3-k_1)^2\,,& \nn\\
&   \cD_7=-n_1 \cdot  k_1\,, 
& &\cD_8=-n_1 \cdot  k_2\,, 
& & \cD_9= -n_1 \cdot k_3\,, & \nn\\
&   \cD_{10}=-n_2 \cdot k_1\,,
& &\cD_{11}=-n_2 \cdot  k_2\,,
& &\cD_{12}=-n_2 \cdot k_3\,, & \nn\\
&    \cD_{13} =-n_J \cdot  k_1 - \omega\,,
& &\cD_{14}=-n_J \cdot  k_2 - \omega\,,
& &\cD_{15}=-n_J \cdot  k_3 - \omega\,,&
\end{align} 
where $\cD_i \to \cD_i -\ri0$ in \eq{Gintegral} is understood.
Our definitions of propagator denominators $\cD_i$ for the remaining ten integral families are given in \app{deftopo}. 
For all eleven families, the result of the generic integral in \eq{Gintegral} is constrained by the scaling properties of its integrand w.r.t.\ \eq{rescalings} to be of the form%
\footnote{Note that $S(\omega)$ vanishes for negative $\omega$.
In view of \eqs{SdefDisc}{TakeDisc} it is therefore natural to pull out a factor of $(-\omega)$, instead of $(+\omega)$, to the power determined by dimensional counting.}
\begin{align}\label{eq:Gres}
G(\vec{a}, \vec{b},\vec{c}, \vec{e},\epsilon)  ={}&
(-\omega-\ri 0)^{3d - 2A - B - C - E}
\left(\frac{n_{12}}{2}\right)^{\frac{3d}{2}-A-B-C}
\left(\frac{n_{1J}}{2}\right)^{A+C-\frac{3d}{2}}
\left(\frac{n_{2J}}{2}\right)^{A+B-\frac{3d}{2}} \nn\\
&\times I(\vec a, \vec b,\vec c, \vec e, \epsilon)\,,
\end{align}
where $A=\sum_i a_i$, $B=\sum_i b_i$, $C=\sum_i c_i$ and $E=\sum_i e_i$. 
The dependence on the external kinematics is thus completely fixed and factored out. What is left to be computed is the dimensionless, in general complex, function $I(\vec a, \vec b,\vec c, \vec e, \epsilon)$. 
In fact, because of \eq{SdefDisc}, we can safely set $\omega=-1$ and only evaluate the real parts of the relevant MIs as an expansion in $\eps$. The $\omega$ dependence of the MIs can finally be restored according to \eq{Gres}.

For the analytic computation of the MIs we follow the approach of \rcite{Bruser:2018rad}, which was inspired by \rcites{Panzer:2014gra,vonManteuffel:2014qoa,vonManteuffel:2015gxa}.
See also  \rcites{Bruser:2019yjk,Liu:2020ydl} for recent applications and some more details of this method.
The basic strategy is to express each MI in terms of known integrals (with less propagators) and integrals that are quasi-finite in $d=4-2\epsilon$ or higher, in practice $d=6-2\epsilon$, $d=8-2\epsilon$, or $d=10-2\epsilon$ dimensions using dimensional recurrence relations \cite{Tarasov:1996br,Lee:2009dh,Lee:2010wea} and IBP reduction.
The integrands of quasi-finite integrals in the Feynman parameter representation are by definition free of endpoint singularities.
For each MI we determine a suitable set of related quasi-finite integrals using the program $\texttt{Reduze2}$~\cite{vonManteuffel:2012np}.
We then expand their integrands and perform the integrations over the Feynman parameters order-by-order in $\epsilon$ using the $\texttt{Maple}$ package $\texttt{HyperInt}$~\cite{Panzer:2014caa}.
In order to deal with threshold singularities (related to the `physical thresholds' discussed in \sec{disc}) $\texttt{HyperInt}$ automatically performs a contour deformation by adding an infinitesimal imaginary part 
to one of the Feynman parameters. 
In general, this procedure leads to unreliable results for the imaginary parts of the MIs (for $\omega=-1$), which is however irrelevant for our calculation, because only the real parts contribute to our soft function.
We discuss the evaluation of the three-loop integrals with non-zero imaginary parts in more detail in \app{HyperIntExample}.
To determine the required dimensional recurrence relations we use $\texttt{LiteRed}$ \cite{Lee:2012cn,Lee:2013mka}. 
Via IBP reduction, we can bring the linear relations between the MIs (collected in the vector $\vec{I}$\,) in $d+2$ and $d$ dimensions into the form $\vec{I}(d+2)={\bm A}(d)\cdot \vec{I}(d)$ for arbitrary $d$. 
We then obtain the analytic results for the MIs by solving the set of linear equations generated by (up to three) iterations of the real matrix ${\bm A}(d)$ with our results for the quasi-finite integrals (in $d=6-2\epsilon$, $d=8-2\epsilon$, and  $d=10-2\epsilon$ dimensions) as the input.
Finally, we check the analytic expressions numerically using the sector decomposition program $\texttt{FIESTA4}$~\cite{Smirnov:2015mct}.


\section{Results}\label{sec:renorm}

Evaluating all relevant three-loop Feynman diagrams as described in the previous section and summing up their contributions yields $\sigma(\omega)$ in \eq{sigmaT} at $\ord{\alpha_s^3}$.
According to \eq{SdefDisc} and using \eq{TakeDisc} we obtain the bare N$^3$LO soft function%
\footnote{Here and in the following we assume $\omega>0$ and do not write out the $\theta(\omega)$ from \eq{TakeDisc} for brevity.}
\begin{align}
S_\eta^{\bare} (\omega) ={}&\delta(\omega)+\frac{Z_\alpha \alpha_s}{4\pi}\frac{1}{\omega}\left(\frac{\mu}{\hat{\omega}}\right)^{2\epsilon}
C_{S_\eta} K_S
\nn\\
&
+\left(\frac{Z_\alpha\alpha_s}{4\pi}\right)^2\frac{1}{\omega}\left(\frac{\mu}{\hat{\omega}}\right)^{4\epsilon}
\left(C_{S_\eta}^2 K_{SS} + C_{S_\eta} C_A K_{SA} +C_{S_\eta} n_f T_F K_{Sf}\right) \nn\\
&+\left(\frac{Z_\alpha\alpha_s}{4\pi}\right)^3
\frac{1}{\omega}\left(\frac{\mu}{\hat{\omega}}\right)^{6\epsilon}
\Bigg[C_{S_\eta}^3 K_{SSS} + C_{S_\eta}^2 C_A K_{SSA} +C_{S_\eta} C_A^2 K_{SAA}
\nn\\
&\qquad\qquad
+C_{S_\eta}^2n_fT_F K_{SSf}
+C_{S_\eta} C_F n_fT_F K_{SFf}+C_{S_\eta} C_A n_fT_F K_{SAf}
\nn\\
&\qquad\qquad
+C_{S_\eta} (n_fT_F)^2 K_{Sff} +\mathcal{C}^{\eta}_3 K_{\mathcal{C}_3}\Bigg] 
+ \cO(\alpha_s^4)\,,
\label{eq:baresmom}
\end{align}
where the coefficients $K_X$ are given in \app{baredata}. 
The subscript $\eta = q\bar q,\,qg,\,gg$ indicates the partonic (production) channel. 
All dependence of the soft function on the kinematic variables $n_{12}$, $n_{1J}$ and $n_{2J}$ is encoded in
\begin{equation}
\hat{\omega} \equiv\omega\sqrt{\frac{2n_{12}}{n_{1J}n_{2J}}}\,,
\end{equation}
which is invariant under the rescaling in \eq{rescalings}.
This follows from the (re)scaling invariance of the Wilson lines in \eq{softfct1}.
Following \rcite{Becher:2012za} we have included a factor of $\sqrt{2}$ in the definition of $\hat{\omega}$ for convenience.

The color factor $C_{S_\eta}$ depends on the partonic channel and can be expressed in terms of the quadratic Casimir invariants $C_F=(N_c^2-1)/(2N_c)$ and $C_A=N_c$ of $SU(N_c)$ with $T_F=1/2$~\cite{Becher:2012za}:
\begin{equation}\label{eq:csdef}
C_{S_{q\bar q}} = C_F-\frac{C_A}{2}\,,
\qquad
C_{S_{qg}} = C_{S_{gg}}=\frac{C_A}{2}\,.
\end{equation}
The color factor $\mathcal{C}^\eta_3$ originates from tripole color structures that first appear at three loops and is computed below.
The number of light (massless) quark flavors is denoted by $n_f$.
For practical reasons our code performing the color algebra for the three-loop Feynman diagrams returns the color factors expressed in terms of $N_c$. 
Due to non-Abelian exponentiation~\cite{Gardi:2010rn,Gardi:2013ita} 
and Casimir scaling of the color dipole contributions 
we are nevertheless able to uniquely reconstruct the color factors in \eq{baresmom} in terms of Casimir invariants as explained in \subsec{renres}.

We define $\alpha_s\equiv \alpha_s(\mu)$ to be the renormalized QCD coupling constant. 
We stress that $Z_\alpha \mu^{2\eps}\alpha_s$ ($=\alpha_s^{\rm bare}$) and therefore the bare soft function is independent of the renormalization scale $\mu$.
For renormalized quantities we use the $\overline{\rm MS}$ scheme throughout this work. The relevant terms of the strong coupling renormalization factor $Z_\alpha$ are
\begin{equation}
Z_\alpha=1+\frac{\alpha_s}{4\pi}\left(-\frac{\beta_0}{\epsilon}\right)
+\left(\frac{\alpha_s}{4\pi}\right)^2\left(\frac{\beta_0^2}{\epsilon^2}
-\frac{\beta_1}{2\epsilon}\right)
+{\cal O}\left(\alpha_s^3\right)
\label{eq:Zalpha}
\end{equation}
with
\begin{equation}
\beta_0= \frac{11}{3}C_A-\frac{4}{3}T_Fn_f  \,,
\qquad 
\beta_1= \frac{34}{3}C_A^2-\frac{20}{3}C_A n_f T_F - 4 C_F n_f T_F \,.
\end{equation}

Following \rcite{Becher:2009th} we conveniently perform the renormalization of the soft function in Laplace space. 
This is, because the Laplace transformation according to
\begin{equation}
\tilde{s}_\eta\left(L\right)=\int_0^\infty  \df \omega \,
\exp\Bigl(-\frac{\omega}{\kappa e^{\gamma_E}}\Bigr) \,S_\eta(\omega)\,,
\label{eq:Laplaces}
\end{equation}
with
\begin{equation}
L=\ln\left(\frac{\kappa}{\mu}\sqrt{\frac{2n_{12}}{n_{1J}n_{2J}}}\right)\,
\end{equation}
and $\kappa$ as defined in \eq{varslaplace}, turns the momentum space convolutions between renormalization factor, anomalous dimension, and renormalized soft function into simple (local) products, cf.\ \eq{xchadthreslap}.
The Laplace transform of the bare momentum space soft function in \eq{baresmom} is obtained using the simple replacement rule
\begin{equation}
\frac{1}{\omega}\left(\frac{\mu}{\hat \omega}\right)^{n\epsilon}\to 
e^{-n\epsilon(L+\gamma_E)}\Gamma(-n\epsilon)\,.
\end{equation}

\subsection{Anomalous dimension}\label{subsec:anomdim}

In Laplace space, the bare and renormalized soft function are related by the local renormalization factor $Z_{s_\eta}$,
\begin{equation}\label{eq:renormaliationdef}
\tilde{s}_\eta^{\bare}(\kappa) = Z_{s_\eta}(\mu) \,\tilde{s}_\eta(L,\mu)\,.
\end{equation}
The renormalized soft function obeys the RGE
\begin{equation}\label{eq:RGequ}
\frac{\rd}{\rd\ln\mu}\tilde{s}_\eta(L,\mu)= \Gamma^{S_\eta}\ \tilde{s}_\eta(L,\mu)
\,,
\end{equation}
with the soft anomalous dimension $\Gamma^{S_\eta}$.
The Laplace space RGEs for the PDFs, the hard, and the jet function take the same form. 
RG invariance of the near-threshold cross section in \eq{xchadthreslap} implies
($\{a,b,c\} = \{q,\bar{q},g\}$, $\{q,g,q\}$, or $\{\bar{q},g,\bar{q}\}$, or $\{g,g,g\}$, $\eta=ab$)
\begin{equation}\label{eq:GammaRGE}
\Gamma^{H_\eta}+ \Gamma^{f_a}+\Gamma^{f_b} + \Gamma^{J_c} +\Gamma^{S_\eta} = 0\,, 
\end{equation}
where%
\footnote{In this paper we use the same convention for $\gamma^{J_i}$ as \rcites{Becher:2006mr,Becher:2009th}. 
In order to switch to the convention of \rcites{Bruser:2018rad,Bruser:2019yjk} 
one has to multiply our $\gamma^{J_i}$ by  $-1/2$.} 
\begin{align}\label{eq:gammafJ}
\Gamma^{f_{i}}={}&2\Gamma_{\rm cusp}^i(\alpha_s) \ln \tau_i
+ 2 \gamma^{f_i}(\alpha_s)\,,\\
\Gamma^{J_{i}}={}&-2\Gamma_{\rm cusp}^i(\alpha_s) \ln \left(\frac{Q^2}{\mu^2}\right) 
- 2 \gamma^{J_i}(\alpha_s)\,,
\end{align}
are the ($x \to 1$ threshold) PDF and jet function anomalous dimensions in Laplace space, 
respectively. 
The variables $Q^2$ and $\tau_i$ are given in \eq{varslaplace} 
and we conveniently define
$\Gamma_{\rm cusp}^q=C_F\gamma_{\rm cusp}$ and $\Gamma_{\rm cusp}^g=C_A\gamma_{\rm cusp}$ for the universal cusp anomalous dimension.

The anomalous dimension of the hard function is denoted by $\Gamma^{H_\eta}$. 
For scattering amplitudes with three massless partons the corresponding operator in  color space reads~\cite{Becher:2009qa,Becher:2019avh}
\begin{equation}
\bm{\Gamma}^H=2\,{\rm Re}\Bigg[\sum_{(i,j)}\frac{\bm{T}_i \cdot \bm{T}_j}{2}
\gcusp(\alpha_s) \ln\frac{\mu^2}{-s_{ij}}+\sum_{i}\gamma^i(\alpha_s) \,\bbid
+f(\alpha_s)\sum_{(i,j,k)}{\cal T}_{iijk}
\Biggr],
\label{eq:gammaH}
\end{equation}
where $\bm{T}_i$ is the color charge operator for the $i$-th parton leg of the scattering amplitude defined in \eq{ColorChargeOps}
and $s_{ij}\equiv 2\sigma_{ij}p_i\cdot p_j+i0$, where the sign factor
 $\sigma_{ij}=+1$, if the momenta of $i$-th and $j$-th parton $p_i$ and $p_j$ are both incoming or outgoing, and $\sigma_{ij}=-1$ otherwise.  
The sums are over all combinations of distinct parton indices. 
The non-cusp anomalous dimension $\gamma^i$ is associated with each external massless (anti)quark ($\gamma^i=\gamma^q$) or gluon ($\gamma^i=\gamma^g$). 
The last term in \eq{gammaH} corresponds to a color tripole contribution and is non-zero starting from three loops. 
The color structure ${\cal T}_{ijkl}$ is given by~\cite{Becher:2009qa}
\begin{equation}
{\cal T}_{iijk} = \frac{1}{2} f^{ade}f^{bce} \{\bm{T}_i^a,\bm{T}_i^b\}\bm{T}_{j}^c\bm{T}_{k}^d \,.
\end{equation}
The three-loop contribution to the associated coefficient
\begin{equation}
f(\alpha_s)=\sum_{n\ge 3}\left(\frac{\alpha_s}{4\pi}\right)^n f_n\,
\end{equation}
was computed in \rcite{Almelid:2015jia} and reads
\begin{equation}
f_3=16(2\zeta_2\zeta_3+\zeta_5)\,.
\end{equation}
Evaluating
\begin{equation}
\Gamma^{H_\eta} = \frac{\bra{a_1a_2a_J}  \bm{\Gamma}^H \ket{a_1a_2a_J}}{\big\langle a_1a_2a_J \big\rvert a_1a_2a_J \big\rangle}
\end{equation}
with
\begin{equation}
\mathcal{C}_3^\eta
\equiv \frac{\bra{a_1a_2a_J} {\cal T}_{iijk}  \ket{a_1a_2a_J}}{\big\langle a_1a_2a_J \big\rvert a_1a_2a_J \big\rangle}
\end{equation}
we obtain from \eq{GammaRGE}
\begin{equation}\label{eq:GammaS}
\Gamma^{S_\eta}
= - 4\Gamma_{\rm cusp}^{\eta}(\alpha_s) \, L
- 2\gamma^{S_\eta}(\alpha_s) - 2 \mathcal{C}_3^{\eta} f(\alpha_s)\,,
\end{equation}
where $\gamma^{S_\eta}$ is the non-cusp piece of the soft anomalous dimension,
and we have
\begin{align}
&\Gamma^{q\bar{q}}_{\rm cusp}=C_{S_{q\bar q}}\gamma_{\rm cusp}\,, 
&&
\gamma^{S_{q\bar{q}}}=2\gamma^q+\gamma^g+2\gamma^{f_q}-\gamma^{J_g}\,, 
&&
\mathcal{C}_3^{q\bar q}
=-6\frac{C_4(F,A)}{C_F} + \frac{C_A^3}{4}
=-\frac{3}{2}N_c\,, 
&\nn\\
&\Gamma^{qg}_{\rm cusp}=C_{S_{qg}}\gamma_{\rm cusp}\,, 
&&
\gamma^{S_{qg}}=2\gamma^q+\gamma^g+\gamma^{f_q}+\gamma^{f_g}-\gamma^{J_q}\,,
&&
\mathcal{C}_3^{qg}
=-6\frac{C_4(F,A)}{C_F} + \frac{C_A^3}{4}
=-\frac{3}{2}N_c\,, 
&\nn\\
&\Gamma^{gg}_{\rm cusp}=C_{S_{gg}}\gamma_{\rm cusp}\,, 
&&
\gamma^{S_{gg}}=3\gamma^g+2\gamma^{f_g}-\gamma^{J_g}\,,
&&
\mathcal{C}^{gg}_3
=-6\frac{C_4(A,A)}{C_A} + \frac{C_A^3}{4}
=-9N_c\,,&
\label{eq:anomaldim}
\end{align}
for the different parton channels.
In order to obtain the explicit expressions for the color factor $\mathcal{C}_3^\eta$ in terms of the quartic Casimir invariants
\begin{equation}
C_4(R,A) = \frac{d_R^{abcd} d_A^{abcd}}{N_R}\,,
\end{equation}
where $d_R^{abcd} = \tr_R\bigl[T_R^{(a} T_R^{b\vphantom{(}} T_R^{c\vphantom{(}} T_R^{d)}\bigr]$ denotes the fully symmetric rank-four tensor of the $SU(N_c)$ representation $R$, and $N_A=N_c^2-1$, $N_F=N_c$, we employed the \texttt{color} code~\cite{vanRitbergen:1998pn}.
All anomalous dimensions on the right-hand sides of \eq{anomaldim} are known at least to three-loop order, see \app{anomdims}.%
\footnote{Starting from four loops the cusp and non-cusp anomalous dimensions violate Casimir scaling~\cite{Lee:2019zop,Henn:2019rmi,Moch:2018wjh}.}
Just like the cusp also the non-cusp anomalous dimension $\gamma^{S_\eta}$ obeys Casimir scaling up to three loops, i.e.\
\begin{equation}\label{eq:CasimirInv}
\frac{\gamma^{S_{q\bar{q}}}}{C_{S_{q\bar{q}}}}
=\frac{\gamma^{S_{qg}}}{C_{S_{qg}}}
=\frac{\gamma^{S_{gg}}}{C_{S_{gg}}}
\,.
\end{equation}

Knowing $\Gamma^{S_\eta}$ we can determine the soft function
renormalization factor $Z_{s_\eta}$ via $\Gamma^{S_\eta}=- \df\ln Z_{s_\eta}/\df\ln\mu$.
Up to three loops we find~\cite{Becher:2009cu,Becher:2009qa}%
\footnote{for brevity we suppress the subscript $\eta$ of the soft function in following.}
\begin{align}
Z_s ={}& 1 + \frac{\alpha_s}{4\pi} \!
\left( \frac{\Gamma^{S\prime}_0}{4\epsilon^2}
+ \frac{\Gamma^S_0}{2\epsilon} \right) 
+ \left( \frac{\alpha_s}{4\pi} \right)^2 \! \left[
\frac{(\Gamma^{S\prime}_0)^2}{32\epsilon^4} 
+ \frac{\Gamma^{S\prime}_0}{8\epsilon^3} 
\bigl( \Gamma^S_0 - \frac32\,\beta_0 \bigr) 
+ \frac{\Gamma^S_0}{8\epsilon^2} 
\left( \Gamma^S_0 -2\beta_0 \right) 
+ \frac{\Gamma^{S\prime}_1}{16\epsilon^2}
+ \frac{\Gamma^S_1}{4\epsilon} \right] \nn\\
&+ \left( \frac{\alpha_s}{4\pi} \right)^3 \Bigg[
\frac{(\Gamma^{S\prime}_0)^3}{384\epsilon^6}
+ \frac{(\Gamma^{S\prime}_0)^2}{64\epsilon^5}
\bigl( \Gamma^S_0 - 3\beta_0 \bigr)
+ \frac{\Gamma^{S\prime}_0}{32\epsilon^4}
\bigl( \Gamma^S_0 - \frac{4}{3}\,\beta_0 \bigr)
\bigl( \Gamma^S_0 - \frac{11}{3}\,\beta_0 \bigr)
+ \frac{\Gamma^{S\prime}_0\Gamma^{S\prime}_1}{64\epsilon^4} \nn\\
&\quad + \frac{\Gamma^S_0}{48\epsilon^3}
\bigl( \Gamma^S_0 - 2\beta_0 \bigr)
\bigl( \Gamma^S_0 - 4\beta_0 \bigr)
+ \frac{\Gamma^{S\prime}_0}{16\epsilon^3}
\bigl( \Gamma^S_1 - \frac{16}{9}\,\beta_1 \bigr)
+ \frac{\Gamma^{S\prime}_1}{32\epsilon^3}
\bigl( \Gamma^S_0 - \frac{20}{9}\,\beta_0 \bigr) \nn\\
&\quad + 
\frac{\Gamma^S_0\Gamma^S_1}{8\epsilon^2} 
- \frac{\beta_0\Gamma^S_1+\beta_1\Gamma^S_0}{6\epsilon^2}
+ \frac{\Gamma^{S\prime}_2}{36\epsilon^2}
+ \frac{\Gamma^S_2}{6\epsilon} \Bigg] 
+ {\cal O}(\alpha_s^4) \,,
\label{eq:ZsNNNLO}
\end{align}
where
\begin{equation}
\Gamma^{S\prime}  \equiv \frac{\partial}{\partial \ln\mu}\Gamma^{S}\,,
\end{equation}
and we have expanded the anomalous dimensions as
\begin{equation}\label{eq:Gbexp}
\Gamma^S = \sum_{n=0}^\infty\,\Gamma^S _n 
\left( \frac{\alpha_s}{4\pi} \right)^{n+1} , \quad
\Gamma^{S\prime} = \sum_{n=0}^\infty\,\Gamma^{S\prime}_n 
\left( \frac{\alpha_s}{4\pi} \right)^{n+1}\,.
\end{equation}
As expected, $Z_{s}$ absorbs all divergences of our explicit result for the bare Laplace space soft function $\tilde{s}(\kappa)$. 
This represents a strong check of our calculation and at the same time confirms the universal infrared structure of QCD scattering amplitudes with three massless parton legs, as predicted by \eq{gammaH}, at three loops.

\subsection{Renormalized results}\label{subsec:renres}

As mentioned above our results for the bare soft function in the different parton channels are initially expressed in terms of $N_c$.
In order to rewrite the color factors in terms of (quadratic and quartic) Casimir invariants we proceed as follows.
Non-Abelian exponentiation~\cite{Gardi:2010rn,Gardi:2013ita}
restricts the three-loop contribution to our soft function for all parton channels to be a universal linear combination of the color factors
\begin{align}
C_S^3\,,\quad
C_S^2 C_A\,,\quad
C_S C_A^2\,,\quad
C_S^2 n_f T_F\,,\quad
C_S C_F n_f T_F\,,\quad
C_S C_A n_f T_F\,,\quad
C_S n_f^2 T_F^2\,,\quad
\mathcal{C}_3\,.
\label{eq:colorfacs}
\end{align}
First, we discuss how to determine the coefficients of the color factors for $n_f=0$.
For the $gg\to g$ channel we have according to \eqs{csdef}{anomaldim} $C_S=N_c/2$ and $\mathcal{C}_3=-9N_c$.
So, all color factors except for $\mathcal{C}_3$ are proportional to $N_c^3$.
We can therefore directly identify the color tripole contribution $\propto \mathcal{C}_3$ in our soft function result.
Once this is fixed we take our result for the $q\bar{q}\to g$ channel and subtract the tripole term with $\mathcal{C}_3^{gg} \to \mathcal{C}_3^{q\bar q}$.
The remaining terms in $S^{\rm bare}_{q\bar q}$ can now be uniquely expressed in terms of the dipole-type color factors in \eq{colorfacs} using the replacement rules
\begin{align}
\frac{1}{N_c^3}\to -8C_{S_{q\bar{q}}}^3\,, \qquad
\frac{1}{N_c}\to4 C_{S_{q\bar{q}}}^2C_A \,, \qquad
N_c\to -2 C_{S_{q\bar{q}}} C_A^2\,.
\end{align}
Concerning the color factors involving $n_f$, we first map the $C_S C_F n_f T_F$ contribution to the $n_f T_F N_c^0$ term in our result for $S^{\rm bare}_{gg}$.
After that we can rewrite the remaining $n_f$ dependent terms in 
$S^{\rm bare}_{q\bar{q}}$ by replacing
\begin{align}
\frac{ n_fT_F}{N_c^2}\to 4 C_{S_{q\bar{q}}}^2 n_fT_F\,, \qquad
n_fT_F N_c^0\to  -2C_{S_{q\bar{q}}} C_A n_fT_F \,, \qquad
 \frac{n_f^2T_F^2}{N_c}\to -2 C_{S_{q\bar{q}}} n_f^2 T_F^2\,.
\end{align}
In this way $S^{\rm bare}_{q\bar q}$ is completely expressed in terms of Casimir invariants and matches \eq{baresmom}. 
We successfully checked the universality of this expression against our explicit results in terms of $N_c$ for $S^{\rm bare}_{qg}$ and $S^{\rm bare}_{gg}$ by adjusting $C_S$ and $\mathcal{C}_3$ according to \eqs{csdef}{anomaldim}.

Solving the RGE in \eq{RGequ} with the anomalous dimension in \eq{GammaS}, the renormalized soft function in Laplace space can be written as
\begin{align}
\tilde{s}(L,\mu)={}&1+\frac{\alpha_s}{4\pi}\left[2\Gamma_0 L^2+2\gamma^S_0 L +c^S_1\right]
+\left(\frac{\alpha_s}{4\pi}\right)^2\Bigg[2 \Gamma _0^2 L^4 -\frac{4}{3} \Gamma _0 \left(\beta _0-3 \gamma _0^S\right)L^3
\nn\\
&\qquad+2 \left(\Gamma _0 c_1^S+\Gamma_1-\beta _0 \gamma _0^S+\left(\gamma_0^{S}\right)^2\right)L^2
+2 \left(c_1^S \left(\gamma _0^S-\beta _0\right)+\gamma_1^S\right)L + c^S_2\Bigg]
\nn\\
&+\left(\frac{\alpha_s}{4\pi}\right)^3
\Bigg[\frac{4 \Gamma_0^3}{3} L^6 + \frac{4\Gamma_0^2}{3} \left(3 \gamma_0^S-2 \beta_0\right)L^5
+\frac{2 \Gamma_0}{3}\Bigg(2 \beta _0^2+3 \Gamma _0 c_1^S+6\Gamma_1+6 \left(\gamma_0^{S}\right)^2
\nn\\
&\qquad\qquad\quad
-10 \beta_0 \gamma_0^S\Bigg)L^4
+\frac{4}{3} \Bigg(3 \gamma _0^S \left(\Gamma _0 c_1^S+\Gamma _1\right)-\beta _0 \left(4 \Gamma _0 c_1^S+2 \Gamma_1+3 \left(\gamma _0^{S}\right)^2\right)
\nn\\
&\qquad\qquad\quad
-\Gamma _0 \left(\beta _1-3 \gamma_1^S\right)+2 \beta _0^2 \gamma _0^S+\left(\gamma _0^{S}\right)^3\Bigg)L^3
+2 \Bigg(\Gamma _0 c_2^S+\Gamma _2 +2 \gamma _0^S \gamma_1^S
\nn\\
&\qquad\qquad\quad
+ c_1^S \left(2 \beta _0^2+\Gamma _1-3 \beta _0 \gamma_0^S+\left(\gamma _0^{S}\right)^2\right)   -\beta_1 \gamma _0^S-2 \beta _0 \gamma _1^S\Bigg)L^2 
\nn\\
&\qquad\qquad\quad
+2 \Bigg(c_1^S \left(\gamma _1^S-\beta _1\right)+c_2^S\left(\gamma _0^S-2 \beta _0\right)+\gamma _2^S + \mathcal{C}_3 f_3\Bigg)L
+c^S_3\Bigg].
\label{eq:finalrensoft}
\end{align}
Here we have expanded the anomalous dimensions in \eq{anomaldim} as
\begin{equation}\label{eq:Ggexp}
\Gamma_{\rm cusp} = \sum_{n=0}^\infty\,\Gamma_n 
\left( \frac{\alpha_s}{4\pi} \right)^{n+1} , \quad
\gamma^{S} = \sum_{n=0}^\infty\,\gamma^{S}_n 
\left( \frac{\alpha_s}{4\pi} \right)^{n+1}\,.
\end{equation}
Our calculation determines the non-logarithmic terms in \eq{finalrensoft} to be
\begin{align}\label{eq:c3general}
c^S_1={}&C_S\pi ^2
\,,\nn\\
c^S_2={}&C_S^2 \frac{\pi ^4}{2}+C_SC_A\left(-\frac{22 \zeta _3}{9}-\frac{14\pi ^4}{15}+\frac{670\pi ^2}{108}+\frac{2428}{81}\right) + C_S n_f T_F\left(\frac{8 \zeta _3}{9}-\frac{50 \pi^2}{27}-\frac{656}{81}\right) 
,\nn\\
c_3^S={}&C_S^3\frac{\pi^6}{6}+C_S^2C_A\Bigg(-\frac{22}{9} \pi ^2 \zeta _3-\frac{14 \pi ^6}{15}+\frac{335 \pi ^4}{54}+\frac{2428 \pi ^2}{81}\Bigg) 
+C_S C_A^2\Bigg(\frac{1108 \zeta_3^2}{9}-\frac{242 \pi ^2 \zeta_3}{27}
\nn\\
&\qquad
-\frac{87052 \zeta_3}{243}-\frac{572 \zeta_5}{9}+\frac{65333 \pi^6}{51030}-\frac{26153\pi^4}{2430}+\frac{256739 \pi^2}{4374}+\frac{5211949}{13122}\Bigg)
\nn\\
&+C_S C_F n_f T_F\Bigg(\frac{112 \pi ^2 \zeta _3}{9}+\frac{5680 \zeta _3}{81}+\frac{448 \zeta _5}{9}+\frac{152 \pi ^4}{405}-\frac{385 \pi ^2}{27}-\frac{42727}{243}\Bigg)
\nn\\
&+C_S C_A n_f T_F \Bigg(-\frac{248}{27} \pi ^2 \zeta _3+\frac{2432 \zeta _3}{81}-\frac{80 \zeta _5}{3}+\frac{464 \pi ^4}{243}-\frac{69254 \pi ^2}{2187}-\frac{825530}{6561}\Bigg)
\nn\\
&+C_S^2 n_f T_F\Bigg(\frac{8 \pi ^2 \zeta _3}{9}-\frac{50 \pi ^4}{27}-\frac{656 \pi ^2}{81}\Bigg)
+C_S n_f^2 T_F^2 \Bigg(\frac{3520 \zeta _3}{243}+\frac{88 \pi ^4}{243}+\frac{784 \pi ^2}{243}-\frac{1024}{6561}\Bigg)
\nn\\
&+2 \mathcal{C}_3  \left(\frac{40 \zeta_3^2}{3}-\frac{1031 \pi ^6}{5670}\right),
\end{align}
The expression for $c_3^S$ is new and represents the main result of this work.

For completeness we also give here the renormalized soft function in momentum space 
\begin{equation}\label{eq:defsmom}
S(\omega,\mu)=\sum_{m=0}^{\infty}\left(\frac{\alpha_s}{4\pi}\right)^m S^{(m)}(\omega,\mu)
\end{equation}
obtained from inverting \eq{Laplaces}. The coefficients in the $\alpha_s$ expansion
take the form
\begin{align}
S^{(m)}(\omega,\mu)=S^{(m)}_{-1}\delta(\omega)+\sum_{n=0}^{2m-1}S^{(m)}_{n}
{\hat \cL}_n\left(\omega\right)\,,
\end{align}
with the plus distributions
\begin{equation}
{\hat \cL}_n\left(\omega\right)
=\left[\frac{\theta(\omega)}{\omega}\ln^n\left(\frac{\hat\omega}{\mu}\right)\right]_+
=\; \lim_{\varepsilon \to 0} \,\frac{\df}{\df \omega}\biggl[ \theta(\omega- \varepsilon)\frac{
	\ln^{n+1} \bigl(\frac{\hat{\omega}}{\mu}\bigr)}{n+1} \biggr]\,.
\end{equation}
The constants $S^{(m)}_{n}$ can be expressed in terms of anomalous dimension coefficients and the $c_i^S$ in \eq{c3general}:
\begin{align}\label{eq:coefsmom}
S^{(1)}_{1}={}&   4\Gamma _0 \,,
\nn\\
S^{(1)}_{0}={}&   2\gamma^S_0 \,,
\nn\\
S^{(1)}_{-1}={}&{-}\frac{\pi^2}{3}\Gamma_0 + c_1^S
\,,\nn\\
S^{(2)}_{3}={}&  8 \Gamma _0^2 
\,,\nn\\
S^{(2)}_{2}={}&  4\Gamma _0 \left(3 \gamma _0^S- \beta _0\right),
\nn\\
S^{(2)}_{1}={}&  4 \Gamma _1+ 4 \Gamma _0 \left(c_1^S- \pi ^2\Gamma _0\right)
+4\gamma _0^{S}\left(\gamma _0^{S}- \beta _0\right),
\nn\\
S^{(2)}_{0}={}& 2\Gamma_0\left(8  \zeta _3 \Gamma _0+ \frac{\pi ^2}{3} \beta _0- \pi ^2 \gamma _0^S\right)
+2\gamma _1^S + 2c_1^S \left( \gamma _0^S- \beta _0\right),
\nn\\
S^{(2)}_{-1}={}& \frac{ \pi ^4}{30} \Gamma _0^2-\frac{\pi ^2 }{3}\Gamma _1
+8 \zeta _3 \Gamma _0 \left(\gamma_0^S-\frac{\beta _0 }{3}\right) +\frac{ \pi ^2}{3}  \gamma _0^S\left(\beta _0-\gamma _0^{S} \right)-\frac{ \pi ^2}{3} \Gamma _0 c_1^S+c_2^S
\,,\nn\\
S^{(3)}_{5}={}&  8 \Gamma _0^3
\,,\nn\\
S^{(3)}_{4}={}& 20 \Gamma _0^2 \left(\gamma _0^S-\frac{2}{3} \beta _0\right),
\nn\\
S^{(3)}_{3}={}& 16\Gamma_0\left[ \frac{\beta _0^2}{3} + \Gamma _1
+\gamma _0^{S} \left(\gamma _0^{S}-\frac{5}{3} \beta _0 \right)\right]
+8 \Gamma _0^2 \left(c_1^S-\frac{5\pi ^2}{3} \Gamma _0\right),
\nn\\
S^{(3)}_{2}={}& 
20\Gamma _0^2\left[8\zeta _3 \Gamma _0+\pi^2\left(\frac{2}{3}\beta_0-\gamma_0^S\right)\right]
+4\Gamma _0 \left[3 \gamma _1^S- \beta _1 + c_1^S  \left(3\gamma _0^S-4\beta _0\right)\right]
+4 \left(\gamma _0^{S}\right)^3 \nn\\
&+4\left(\beta _0 \gamma _0^S-\Gamma _1 \right)\left( 2 \beta _0-3 \gamma _0^{S}\right),
\nn\\
S^{(3)}_{1}={}&4 \Gamma _2+  \frac{2\pi ^4}{3} \Gamma _0^3 
+160 \zeta _3 \Gamma _0^2 \left(\gamma _0^S-\frac{2\beta _0}{3}\right)
+8\pi ^2\Gamma _0 \left[-\frac{ \beta _0^2}{3}-\Gamma _1
+\frac{5}{3}  \beta _0 \gamma _0^S- \left(\gamma _0^{S}\right)^2 \right]\nn\\
&-8 \beta _0 \gamma _1^S+4\gamma _0^S\left(2\gamma _1^S- \beta _1\right)
+ 4 c_1^S \left[ \left(2\beta_0-\gamma _0^S\right)\left(\beta_0-\gamma _0^S\right)
- \pi ^2 \Gamma _0^2+\Gamma _1\right]
+4 \Gamma _0 c_2^S 
\,,\nn\\
S^{(3)}_{0}={}&  16\Gamma _0^3 \left(12 \zeta _5-\frac{5 \pi ^2 \zeta _3}{3}\right)
+\frac{\pi ^4 }{3} \Gamma _0^2 \left(\gamma _0^S-\frac{2}{3}\beta_0\right)
+2\pi ^2\Gamma _1 \left(\frac{2}{3} \beta_0-\gamma _0^S\right) +2 \pi^2 \Gamma _0 \left(\frac{\beta _1}{3}-\gamma _1^S\right)
\nn\\
&+32\zeta_3\Gamma _0 \left[\frac{\beta _0^2}{3} + \Gamma _1 
-\frac{5}{3} \beta _0  \gamma _0^S+\left( \gamma _0^{S}\right)^2\right]
+2 \gamma _2^S
-\frac{2\pi^2}{3}\gamma _0^{S}
\left(2\beta _0-\gamma _0^{S}\right)\left(\beta _0-\gamma _0^{S}\right)
\nn\\
&+2c_1^S \left[- \beta _1+8\zeta _3 \Gamma _0^2 
+ \pi ^2\Gamma _0 \left(\frac{4 }{3}\beta _0- \gamma _0^S\right)+ \gamma _1^S\right]
+2c_2^S\left( \gamma _0^S- 2\beta _0\right) + 2\mathcal{C}_3 f_3
\,,\nn\\
S^{(3)}_{-1}={}& \frac{5}{3} \Gamma _0^3 \left(32\zeta _3^2-\frac{\pi ^6}{42}\right)
+\Gamma _0^2\left(\frac{40\pi^2\zeta_3}{3}-96\zeta_5\right)\left(\frac{2}{3}\beta_0-\gamma_0^S\right)
+8 \zeta _3\Gamma _1 \left( \gamma_0^S-\frac{2}{3} \beta _0 \right)
\nn\\
&-\frac{\pi ^2 }{3}\Gamma _2+\Gamma _0 \left[\frac{ \pi ^4 }{45}\left(\beta _0^2-5\beta _0 \gamma _0^S
+3\left(\gamma _0^{S}\right)^2+3\Gamma_1\right)
+8 \zeta _3 \left( \gamma _1^S-\frac{\beta _1}{3} \right)\right]
\nn\\
&
+\frac{8\zeta _3 }{3} \gamma _0^{S}\left(2\beta_0-\gamma_0^S\right)\left(\beta_0-\gamma_0^S\right)
+\frac{\pi ^2}{3} \gamma _0^S \left( \beta _1-2 \gamma _1^S\right)
+\frac{2\pi ^2 }{3} \beta _0 \gamma _1^S
-\frac{\pi ^2}{3}\Gamma _0 c_2^S+c_3^S 
\nn\\
&+c_1^S \left[
\frac{\pi ^4 }{30} \Gamma _0^2-\frac{\pi ^2 }{3}\Gamma _1+8\zeta _3\Gamma _0 \left( \gamma_0^S-\frac{4}{3}\beta _0 \right)
-\frac{\pi ^2}{3} \left( 2\beta_0-\gamma_0^S\right)\left( \beta_0-\gamma_0^S\right)\right].
\end{align}

\section{Conclusion}\label{sec:conclu}

We have calculated the universal three-loop soft function contributing to factorized cross sections for EW boson production in the (threshold) limit, where the transverse momentum of the boson is close to its kinematically allowed maximum for a given (not too large) rapidity.
To the best of our knowledge this represents the first result of a soft function for a three-parton process at three loops.
We have derived a novel expression for the soft function in terms of a forward-scattering-type matrix element of Wilson line operators. This allowed us to avoid any phase space integrations and to straight-forwardly take advantage of well-established multi-loop technology in our calculation.

We present our results for the renormalized soft function both in Laplace space and momentum space in \subsec{renres}. The three-loop non-logarithmic terms in \eq{c3general} represent the genuinely new information at this order. 
Our explicit three-loop calculation validates the relation between the anomalous dimensions of the PDFs and the corresponding hard, jet, and soft functions inferred from RG consistency of the near-threshold cross section.
In this way we also confirm the nonzero (Casimir scaling violating) color tripole contribution to the three-loop soft and corresponding hard anomalous dimensions found in \rcite{Almelid:2015jia}.

The threshold approximation of the EW boson production cross section is strictly valid when the invariant mass of the hadronic radiation (including the proton remnants) is small compared to the transverse momentum of the boson.  
Phenomenologically, however, it usually performs well even far away from this limit, see e.g.\ \rcites{Becher:2011fc,Becher:2012xr}.
Together with the three-loop jet function~\cite{Bruser:2018rad,Banerjee:2018ozf} and the yet-unknown three-loop hard function our soft function result will allow to determine the N$^3$LO corrections to the transverse-momentum spectra of $\gamma$, $W^\pm$, $Z$ and Higgs bosons in the (far-)tail region.
Furthermore, it is necessary to carry out the resummation of threshold logarithms to N$^3$LL$^\prime$ (or N$^4$LL) accuracy.
Once the QCD three-loop virtual corrections to the $q\bar{q} \to Yg$ (or crossing-symmetric) and $gg \to Hg$ scattering amplitudes, i.e.\ the hard functions, become available, 
this may constitute significant progress in the precision phenomenology for Higgs, photon and EW gauge boson production in association with a jet at the LHC.

\begin{acknowledgments}
Z.L.L.\ is grateful to Thomas Becher, Matthias Neubert and Ding Yu Shao for valuable discussions.
M.S.\ thanks Erik Panzer for inspiring feedback and explanations regarding $\texttt{HyperInt}$.
 We are indebted to Robin Br\"user for providing the codes to perform the diagram generation, topology mapping, and partial fractioning and for his help with the $\texttt{color}$ code. This research has been supported by the Cluster of Excellence PRISMA$^+$ (project ID 39083149), funded by the German Research Foundation (DFG). The research of Z.L.L.\ is supported by the U.S. Department of Energy under Contract No.\ DE-AC52-06NA25396, the LANL/LDRD program and within the framework of the TMD Topical Collaboration.
All graphs were drawn using \texttt{JaxoDraw}~\cite{Binosi:2008ig}.
\end{acknowledgments}

\appendix
\section{Three-loop anomalous dimensions}
\label{app:anomdims}
For completeness we list here the explicit expressions for all anomalous dimensions in \eq{anomaldim} to three-loop order. 
The convention for the loop expansions is analogous to \eqs{Gbexp}{Ggexp}. 
The coefficients of the cusp anomalous dimension are~\cite{Korchemsky:1987wg,Moch:2004pa}
\begin{align}
\gamma_0^{\rm cusp} ={}& 4 \,, \\
\gamma_1^{\rm cusp} ={}& \biggr( \frac{268}{9} 
- \frac{4\pi^2}{3} \biggl) C_A - \frac{80}{9}\,T_F n_f \,,
\\
\gamma_2^{\rm cusp} ={}& C_A^2 \left( \frac{490}{3} 
- \frac{536\pi^2}{27}
+ \frac{44\pi^4}{45} + \frac{88}{3}\,\zeta_3 \right) 
+ C_A T_F n_f  \left( - \frac{1672}{27} + \frac{160\pi^2}{27}
- \frac{224}{3}\,\zeta_3 \right) \nonumber\\
&+ C_F T_F n_f \left( - \frac{220}{3} + 64\zeta_3 \right) 
- \frac{64}{27}\,T_F^2 n_f^2 \,.
\end{align}
The coefficients of the soft non-cusp anomalous dimension $\gamma^S$ are~\cite{Becher:2009th}
\begin{align}
\gamma_0^S={}&0 \,,
\\
\gamma_1^S={}&C_S\Bigg[C_A\left(28 \zeta _3+\frac{11 \pi ^2}{9}-\frac{808}{27}\right) +n_f T_F\left(\frac{224}{27}-\frac{4 \pi ^2}{9}\right) 
\Bigg],
\\
\gamma_2^S={}& C_S\Bigg[C_A^2\left(-\frac{88}{9} \pi ^2 \zeta_3+\frac{1316 \zeta _3}{3}-192 \zeta _5-\frac{88 \pi ^4}{45}+\frac{6325 \pi^2}{243}-\frac{136781}{729}\right) 
\nn\\
&\quad+ C_A n_f T_F \left(-\frac{1456 \zeta _3}{27}+\frac{16 \pi ^4}{15}-\frac{2828 \pi^2}{243}+\frac{23684}{729}\right)
\nn\\
&\quad+C_Fn_f T_F \left(\frac{3422}{27}-\frac{4 \pi ^2}{3}-\frac{608\zeta_3}{9}-\frac{16 \pi ^4}{45}\right)
+n_f^2 T_F^2 \left(\frac{8320}{729}+\frac{80 \pi^2}{81}-\frac{448 \zeta _3}{27}\right) 
\Bigg].
\end{align}
The hard anomalous dimensions $\gamma^q$ and $\gamma^g$ can be determined~\cite{Becher:2009qa} from the divergent part of the on-shell quark and gluon form factors in QCD \cite{Moch:2005id,Moch:2005tm} and read up to three loops
\begin{align}
\gamma_0^q ={}& {-}3 C_F \,, \\
\gamma_1^q ={}& C_F^2 \left( -\frac{3}{2} + 2\pi^2
- 24\zeta_3 \right)
+ C_F C_A \left( - \frac{961}{54} - \frac{11\pi^2}{6} 
+ 26\zeta_3 \right)
+ C_F T_F n_f \left( \frac{130}{27} + \frac{2\pi^2}{3} \right),
\\
\gamma_2^q ={}& C_F^3 \left( -\frac{29}{2} - 3\pi^2
- \frac{8\pi^4}{5}
- 68\zeta_3 + \frac{16\pi^2}{3}\,\zeta_3 + 240\zeta_5 \right) 
\nonumber\\
&+ C_F^2 C_A \left( - \frac{151}{4} + \frac{205\pi^2}{9}
+ \frac{247\pi^4}{135} - \frac{844}{3}\,\zeta_3
- \frac{8\pi^2}{3}\,\zeta_3 - 120\zeta_5 \right) \nonumber\\
&+ C_F C_A^2 \left( - \frac{139345}{2916} - \frac{7163\pi^2}{486}
- \frac{83\pi^4}{90} + \frac{3526}{9}\,\zeta_3
- \frac{44\pi^2}{9}\,\zeta_3 - 136\zeta_5 \right) \nonumber\\
&+ C_F^2 T_F n_f \left( \frac{2953}{27} - \frac{26\pi^2}{9} 
- \frac{28\pi^4}{27} + \frac{512}{9}\,\zeta_3 \right) 
\nonumber\\
&+ C_F C_A T_F n_f \left( - \frac{17318}{729}
+ \frac{2594\pi^2}{243} + \frac{22\pi^4}{45} 
- \frac{1928}{27}\,\zeta_3 \right) \nonumber\\
&+ C_F T_F^2 n_f^2 \left( \frac{9668}{729} 
- \frac{40\pi^2}{27} - \frac{32}{27}\,\zeta_3 \right) \,,
\\[2 ex]
\gamma_0^g ={}& {-} \beta_0 
= - \frac{11}{3}\,C_A + \frac43\,T_F n_f \,, \\
\gamma_1^g ={}& C_A^2 \left( -\frac{692}{27} + \frac{11\pi^2}{18}
+ 2\zeta_3 \right) 
+ C_A T_F n_f \left( \frac{256}{27} - \frac{2\pi^2}{9} \right)
+ 4 C_F T_F n_f \,, 
\\
\gamma_2^g ={}& C_A^3 \left( - \frac{97186}{729} 
+ \frac{6109\pi^2}{486} - \frac{319\pi^4}{270} 
+ \frac{122}{3}\,\zeta_3 - \frac{20\pi^2}{9}\,\zeta_3 
- 16\zeta_5 \right) \nonumber\\
&+ C_A^2 T_F n_f \left( \frac{30715}{729}
- \frac{1198\pi^2}{243} + \frac{82\pi^4}{135} 
+ \frac{712}{27}\,\zeta_3 \right) \nonumber\\
&+ C_A C_F T_F n_f \left( \frac{2434}{27} 
- \frac{2\pi^2}{3} - \frac{8\pi^4}{45} 
- \frac{304}{9}\,\zeta_3 \right) 
- 2 C_F^2 T_F n_f \nonumber\\
&+ C_A T_F^2 n_f^2 \left( - \frac{538}{729}
+ \frac{40\pi^2}{81} - \frac{224}{27}\,\zeta_3 \right) 
- \frac{44}{9}\,C_F T_F^2 n_f^2 \,.
\end{align}
At three loops the anomalous dimensions of the quark and gluon jet functions were derived in \rcites{Becher:2006mr,Becher:2009th} based on RG consistency arguments and confirmed by the jet function calculations in \rcites{Bruser:2018rad,Banerjee:2018ozf}.
Their coefficients are
\begin{align}
\gamma_0^{J_q} ={}& {-}3 C_F \,, \\
\gamma_1^{J_q} ={}& C_F^2 \left( - \frac{3}{2} + 2\pi^2 - 24\zeta_3 \right) 
+ C_F C_A \left( - \frac{1769}{54} - \frac{11\pi^2}{9} + 40\zeta_3 \right)
+ C_F T_F n_f \left( \frac{242}{27} + \frac{4\pi^2}{9} \right), 
\\
\gamma_2^{J_q}
={}& C_F^3 \left( - \frac{29}{2} - 3\pi^2 - \frac{8\pi^4}{5} - 68\zeta_3 
+ \frac{16\pi^2}{3}\,\zeta_3 + 240\zeta_5 \right) \nonumber\\
&+ C_F^2 C_A \left( - \frac{151}{4} + \frac{205\pi^2}{9}
+ \frac{247\pi^4}{135} - \frac{844}{3}\,\zeta_3
- \frac{8\pi^2}{3}\,\zeta_3 - 120\zeta_5 \right) \nonumber\\
&+ C_F C_A^2 \left( - \frac{412907}{2916} - \frac{419\pi^2}{243}
- \frac{19\pi^4}{10} + \frac{5500}{9}\,\zeta_3
- \frac{88\pi^2}{9}\,\zeta_3 - 232\zeta_5 \right) \nonumber\\
&+ C_F^2 T_F n_f \left( \frac{4664}{27} - \frac{32\pi^2}{9}
- \frac{164\pi^4}{135} + \frac{208}{9}\,\zeta_3 \right) \nonumber\\
&+ C_F C_A T_F n_f \left( - \frac{5476}{729}
+ \frac{1180\pi^2}{243} + \frac{46\pi^4}{45}
- \frac{2656}{27}\,\zeta_3 \right) \nonumber\\
&+ C_F T_F^2 n_f^2 \left( \frac{13828}{729} - \frac{80\pi^2}{81}
- \frac{256}{27}\,\zeta_3 \right),
\end{align}
and
\begin{align}
\gamma_0^{J_g} ={}&{-} \beta_0 \, ,\\
\gamma_1^{J_g} ={}& C_A^2 \left( - \frac{1096}{27} + \frac{11 \pi^2}{9} + 16  \zeta_3 \right) + C_A n_f T_F \left( \frac{368}{27} - \frac{4 \pi^2}{9}  \right) + 4 C_F T_F n_f \,, \\
\gamma_2^{J_g} ={}& \left( - \frac{331153}{1458} + \frac{6217 \pi^2}{243} + 260
\zeta_3 - \frac{583 \pi^4}{270} - \frac{64 \pi^2 \zeta_3}{9} - 112 \zeta_5
\right) C_A^3  \nn\\
&  + \left( \frac{42557}{729} - \frac{2612}{243} - \frac{16 \zeta_3}{27} +
\frac{154 \pi^4}{135} \right) C_A^2 n_f T_F 
+ \left( \frac{3622}{729} + \frac{80 \pi^2}{81} - \frac{448 \zeta_3}{27}  \right) C_A n_f^2 T_F^2 \nn \\
&  + \left( \frac{4145}{27} - \frac{4 \pi^2}{3} - \frac{608 \zeta_3}{9} -  \frac{16 \pi^4}{45} \right) C_A C_F n_f T_F 
- 2 C_F^2 n_f T_F - \frac{44}{9} C_F n_F^2 T_F^2 \,,
\end{align}
respectively.
The anomalous dimensions describing the evolution of the quark and gluon PDFs near $x\to 1$ can be determined from the QCD splitting functions~\cite{Moch:2004pa,Vogt:2004mw} and were extracted in \rcites{Becher:2007ty,Ahrens:2008nc} up to three-loop order:
\begin{align}
\gamma_0^{f_q} ={}& 3 C_F \,, \\
\gamma_1^{f_q} 
={}& C_F^2 \left( \frac{3}{2} - 2\pi^2 + 24\zeta_3 \right) 
+ C_F C_A \left( \frac{17}{6} + \frac{22\pi^2}{9} 
- 12\zeta_3 \right)
- C_F T_F n_f \left( \frac{2}{3} + \frac{8\pi^2}{9} \right) ,
\\
\gamma_2^{f_q}
={}& C_F^3 \left( \frac{29}{2} + 3\pi^2 + \frac{8\pi^4}{5} 
+ 68\zeta_3 - \frac{16\pi^2}{3}\,\zeta_3 - 240\zeta_5 \right)
\nn\\
&+ C_F^2 C_A \left( \frac{151}{4} - \frac{205\pi^2}{9}
- \frac{247\pi^4}{135} + \frac{844}{3}\,\zeta_3
+ \frac{8\pi^2}{3}\,\zeta_3 + 120\zeta_5 \right) \nn\\
&+ C_F^2 T_F n_f \left( - 46 + \frac{20\pi^2}{9}
+ \frac{116\pi^4}{135} - \frac{272}{3}\,\zeta_3 \right) 
\nn\\
&+ C_F C_A^2 \left( - \frac{1657}{36}
+ \frac{2248\pi^2}{81} - \frac{\pi^4}{18} 
- \frac{1552}{9}\,\zeta_3 + 40\zeta_5 \right) \nn\\
&+ C_F C_A T_F n_f \left( 40 - \frac{1336\pi^2}{81}
+ \frac{2\pi^4}{45} + \frac{400}{9}\,\zeta_3 \right) \nn\\
&+ C_F T_F^2 n_f^2 \left( - \frac{68}{9} 
+ \frac{160\pi^2}{81} - \frac{64}{9}\,\zeta_3 \right) ,
\\[2 ex]
\gamma^{f_g}_0 
={}& \frac{11}{3}\,C_A - \frac{4}{3}\,T_F n_f
= \beta_0 \,, \\
\gamma^{f_g}_1 
={}& 4 C_A^2 \left( \frac{8}{3} + 3\zeta_3 \right) 
- \frac{16}{3}\,C_A T_F n_f - 4 C_F T_F n_f \,, \\
\gamma^{f_g}_2 
={}& C_A^3 \left[ \frac{79}{2} + \frac{4\pi^2}{9} + \frac{11\pi^4}{54} 
+ \left( \frac{536}{3} - \frac{8\pi^2}{3} \right) \zeta_3 
- 80\zeta_5 \right] \nn\\
&- C_A^2 T_F n_f \left( \frac{233}{9} + \frac{8\pi^2}{9}
+ \frac{2\pi^4}{27} + \frac{160}{3}\,\zeta_3 \right) \nn\\
&- \frac{241}{9}\,C_A C_F T_F n_f + 2 C_F^2 T_F n_f
+ \frac{58}{9}\,C_A T_F^2 n_f^2 + \frac{44}{9}\,C_F T_F^2 n_f^2 
\,.
\end{align}

\section{Bare data}\label{app:baredata} 
Here we present our expressions for the coefficients of the different color 
structures in the bare soft function, \eq{baresmom}. 
We show the results as 
an expansion in $\eps = (4-d)/2$ to the order required for the calculation of 
the renormalized three-loop soft function:
\begin{align}
K_S={}&-\frac{8}{\epsilon }+\frac{2 \pi ^2 \epsilon }{3}+\frac{56 \zeta _3 \epsilon ^2}{3}+\frac{47 \pi ^4 \epsilon ^3}{180}+\Bigg(\frac{248 \zeta
	_5}{5}-\frac{14 \pi ^2 \zeta _3}{9}\Bigg) \epsilon ^4+\Bigg(\frac{949 \pi ^6}{15120}-\frac{196 \zeta _3^2}{9}\Bigg) \epsilon ^5
\nn\\
&+\cO\left(\epsilon^6\right),
\\
K_{SS}={}&-\frac{32}{\epsilon ^3}+\frac{80 \pi ^2}{3 \epsilon }+\frac{1984 \zeta _3}{3}+\frac{100 \pi ^4 \epsilon }{9}+\Bigg(\frac{32704 \zeta_5}{5}-\frac{4960 \pi ^2 \zeta _3}{9}\Bigg) \epsilon ^2
\nn\\
&+\Bigg(\frac{482 \pi ^6}{63}-\frac{61504 \zeta _3^2}{9}\Bigg) \epsilon^3+\cO\left(\epsilon^4\right),
\\
K_{SA}={}&-\frac{44}{3 \epsilon ^2}+\frac{1}{\epsilon }\Bigg(\frac{4 \pi ^2}{3}-\frac{268}{9}\Bigg)
+\Bigg(56 \zeta _3+\frac{22 \pi^2}{3}-\frac{1616}{27}\Bigg)
+\Bigg(\frac{2728 \zeta _3}{9}+\frac{88 \pi ^4}{45}+\frac{134 \pi ^2}{9}
\nn\\
&\qquad-\frac{9712}{81}\Bigg) \epsilon
+\Bigg(\frac{16616 \zeta _3}{27}-\frac{716 \pi ^2 \zeta _3}{9}+296 \zeta _5+\frac{649 \pi ^4}{90}+\frac{808 \pi^2}{27}-\frac{58304}{243}\Bigg) \epsilon ^2
\nn\\
&+\Bigg(-\frac{3688 \zeta _3^2}{3}+\frac{100192\zeta_3}{81}-\frac{1364 \pi ^2\zeta_3}{9}+\frac{44968 \zeta _5}{15}-\frac{377 \pi ^6}{1890}+\frac{3953 \pi ^4}{270}+\frac{4856 \pi^2}{81}
\nn\\
&\qquad-\frac{349888}{729}\Bigg) \epsilon^3
+\cO\left(\epsilon^4\right),
\\
K_{Sf}={}&\frac{16}{3 \epsilon ^2}+\frac{80}{9 \epsilon }+\Bigg(\frac{448}{27}-\frac{8 \pi ^2}{3}\Bigg)+\Bigg(-\frac{992 \zeta _3}{9}-\frac{40 \pi^2}{9}+\frac{2624}{81}\Bigg) \epsilon \nn\\
&+\Bigg(-\frac{4960 \zeta _3}{27}-\frac{118 \pi ^4}{45}-\frac{224 \pi^2}{27}+\frac{15616}{243}\Bigg)\epsilon ^2
\nn\\
&+\Bigg(\frac{496 \pi^2\zeta _3}{9}-\frac{27776\zeta _3}{81}-\frac{16352 \zeta _5}{15}-\frac{118 \pi^4}{27}-\frac{1312 \pi^2}{81}+\frac{93440}{729}\Bigg) \epsilon ^3+\cO\left(\epsilon^4\right),
\\
K_{SSS}={}&-\frac{64}{\epsilon ^5}+\frac{144 \pi ^2}{\epsilon ^3}+\frac{4544 \zeta _3}{\epsilon ^2}+\frac{334 \pi ^4}{5 \epsilon }+\Bigg(\frac{497472 \zeta_5}{5}-10224 \pi ^2 \zeta _3\Bigg)
\nn\\
&+\Bigg(\frac{83933 \pi ^6}{630}-161312 \zeta _3^2\Bigg) \epsilon 
+\cO\left(\epsilon^2\right),
\\
K_{SSA}={}&-\frac{88}{\epsilon ^4}+\frac{1}{\epsilon ^3}\Bigg(8 \pi ^2-\frac{536}{3}\Bigg)
+\frac{1}{\epsilon ^2}\Bigg(336 \zeta _3+\frac{506 \pi ^2}{3}-\frac{3232}{9}\Bigg)
\nn\\
&+\frac{1}{\epsilon}\Bigg(6248\zeta _3+\frac{2 \pi ^4}{5}+\frac{3082 \pi^2}{9}-\frac{19424}{27}\Bigg)
\nn\\
&+\Bigg(\frac{38056}{3}\zeta _3-1356 \pi ^2\zeta _3 +1776 \zeta_5+\frac{8767 \pi ^4}{60}+\frac{18584\pi^2}{27}-\frac{116608}{81}\Bigg)
\nn\\
&+\Bigg(-24288\zeta_3^2+\frac{229472}{9}\zeta _3-\frac{35926}{3} \pi^2\zeta _3
+\frac{684024 \zeta _5}{5}-\frac{4597 \pi ^6}{140}+\frac{53399 \pi ^4}{180}
\nn\\
&\qquad
+\frac{111688 \pi^2}{81}-\frac{699776}{243}\Bigg) \epsilon 
+\cO\left(\epsilon^2\right),
\\
K_{SAA}={}&-\frac{968}{27 \epsilon ^3}+\frac{1}{\epsilon ^2}\Bigg(\frac{176 \pi ^2}{27}-\frac{13016}{81}\Bigg)
+\frac{1}{\epsilon}\Bigg(\frac{3520 \zeta _3}{9}-\frac{88 \pi
	^4}{135}+\frac{4702 \pi ^2}{81}-\frac{44372}{81}\Bigg)
\nn\\
&+\Bigg(\frac{10376}{3} \zeta _3-\frac{176 }{9} \pi ^2\zeta _3-384 \zeta
_5+\frac{88 \pi ^4}{5}+\frac{55202 \pi ^2}{243}-\frac{1235050}{729}\Bigg)
\nn\\
&+\Bigg(-\frac{2216 \zeta _3^2}{3}+\frac{1089896}{81}\zeta _3-\frac{11968}{9}\pi ^2\zeta _3 +\frac{10912 \zeta _5}{3}-\frac{48701 \pi ^6}{8505}+\frac{198197 \pi ^4}{1620}
\nn\\
&\qquad
+\frac{541345 \pi^2}{729}-\frac{10984045}{2187}\Bigg) \epsilon 
+\cO\left(\epsilon^2\right),
\\
K_{SSf}={}&\frac{32}{\epsilon ^4}+\frac{160}{3 \epsilon ^3}
+\frac{1}{\epsilon ^2}\Bigg(\frac{896}{9}-\frac{184 \pi ^2}{3}\Bigg)
+\frac{1}{\epsilon }\Bigg(-2272 \zeta _3-\frac{920 \pi^2}{9}+\frac{5248}{27}\Bigg)
\nn\\
&+\Bigg(-\frac{11360 \zeta _3}{3}-\frac{797 \pi ^4}{15}-\frac{5152 \pi^2}{27}+\frac{31232}{81}\Bigg)
\nn\\
&+\Bigg(\frac{13064 \pi ^2\zeta _3}{3}-\frac{63616\zeta _3}{9} -\frac{248736 \zeta _5}{5}-\frac{797 \pi^4}{9}
-\frac{30176 \pi ^2}{81}+\frac{186880}{243}\Bigg) \epsilon 
+\cO\left(\epsilon^2\right),
\\
K_{SFf}={}&\frac{16}{3 \epsilon ^2}
+\frac{1}{\epsilon}\Bigg(\frac{440}{9}-\frac{128 \zeta _3}{3}\Bigg)
+\Bigg(-\frac{1216 \zeta _3}{9}-\frac{32 \pi ^4}{45}-\frac{20 \pi^2}{3}+\frac{6844}{27}\Bigg)
\nn\\
&+\Bigg(\frac{160 \pi^2\zeta _3}{3}-\frac{21296\zeta _3}{27}-\frac{896 \zeta _5}{3}-\frac{304 \pi^4}{135}-\frac{550 \pi ^2}{9}+\frac{85454}{81}\Bigg) \epsilon +\cO\left(\epsilon^2\right),
\\
K_{SAf}={}&\frac{704}{27 \epsilon ^3}
+\frac{1}{\epsilon ^2}\Bigg(\frac{8528}{81}-\frac{64 \pi ^2}{27}\Bigg)
+\frac{1}{\epsilon }\Bigg(-\frac{896 \zeta _3}{9}-\frac{2960 \pi^2}{81}+\frac{26128}{81}\Bigg)
\nn\\
&+\Bigg(-\frac{53600 \zeta _3}{27}-\frac{256 \pi ^4}{45}-\frac{33628 \pi^2}{243}+\frac{656776}{729}\Bigg)
\nn\\
&+\Bigg(\frac{3872 \pi ^2\zeta _3}{9}-\frac{22768\zeta _3}{3}-1024 \zeta _5-\frac{5710 \pi^4}{81}-\frac{310628 \pi^2}{729}+\frac{5295988}{2187}\Bigg) \epsilon 
+\cO\left(\epsilon^2\right),
\\
K_{Sff}={}&-\frac{128}{27 \epsilon ^3}-\frac{1280}{81 \epsilon ^2}
+\frac{1}{\epsilon }\Bigg(\frac{160 \pi ^2}{27}-\frac{128}{3}\Bigg)
+\Bigg(\frac{8320 \zeta_3}{27}+\frac{1600 \pi^2}{81}-\frac{77824}{729}\Bigg)
\nn\\
&+\Bigg(\frac{83200 \zeta _3}{81}+\frac{868 \pi ^4}{81}+\frac{160 \pi^2}{3}-\frac{560128}{2187}\Bigg) \epsilon 
+\cO\left(\epsilon^2\right)
.
\end{align}

\section{Definition of integral families}\label{app:deftopo}
In section~\ref{sec:calc}, we defined one integral family for illustration.
All eleven families share the same propagator denominators $\cD_{1,\cdots,6}$.
Here we present the definition of the other propagator denominators $\cD_{7,\cdots,15}$ for the remaining ten families.
\begin{align}
&\text{Family 2:}
\nn\\
&\quad   \cD_7=-n_1 \!\cdot\!  k_1\,, 
& &\cD_8=-n_1 \!\cdot\!  k_2\,, 
& & \cD_9= -n_1 \!\cdot\! k_3\,, & \nn\\
&\quad   \cD_{10}=-n_2 \!\cdot\! (k_3-k_1)\,,
& &\cD_{11}=-n_2 \!\cdot\! (k_3-k_2)\,,
& &\cD_{12}=-n_2 \!\cdot\! k_3\,, & \\
&\quad    \cD_{13} =-n_J \!\cdot\!  k_1 - \omega\,,
& &\cD_{14}=-n_J \!\cdot\!  k_2 - \omega\,,
& &\cD_{15}=-n_J \!\cdot\!  k_3 - \omega\,.&
\nn\\ 
&\text{Family 3:}
\nn\\
&\quad   \cD_7=-n_1 \!\cdot\!  k_1\,, 
& &\cD_8=-n_1 \!\cdot\!  k_2\,, 
& & \cD_9= -n_1 \!\cdot\! (k_3-k_2)\,, & \nn\\
&\quad   \cD_{10}=-n_2 \!\cdot\! k_1\,,
& &\cD_{11}=-n_2 \!\cdot\!  k_2\,,
& &\cD_{12}=-n_2 \!\cdot\! (k_3-k_2)\,, & \\
&\quad    \cD_{13} =-n_J \!\cdot\!  k_1 - \omega\,,
& &\cD_{14}=-n_J \!\cdot\!  k_2 - \omega\,,
& &\cD_{15}=-n_J \!\cdot\!  k_3 - \omega\,.&
\nn\\ 
&\text{Family 4:}
\nn\\
&\quad    \cD_7=-n_1 \!\cdot\!  k_1\,, 
& &\cD_8=-n_1 \!\cdot\!  (k_3-k_2)\,, 
& & \cD_9= -n_1 \!\cdot\! k_3\,, & \nn\\
&\quad    \cD_{10}=n_2 \!\cdot\! k_1\,,
& &\cD_{11}=-n_2 \!\cdot\! (k_2-k_1) \,,
& &\cD_{12}=-n_2 \!\cdot\! (k_3-k_1)\,, & \\
&\quad     \cD_{13} =-n_J \!\cdot\!  k_3 - \omega\,,
& &\cD_{14}=-n_J \!\cdot\!  (k_3-k_2) - \omega\,,
& &\cD_{15}=-n_J \!\cdot\!  (k_3-k_1)  - \omega\,.&
\nn\\ 
&\text{Family 5:}
\nn\\
&\quad    \cD_7=-n_1 \!\cdot\!  k_1\,, 
& &\cD_8=-n_1 \!\cdot\!  k_2\,, 
& & \cD_9= -n_1 \!\cdot\! k_3\,, & \nn\\
&\quad    \cD_{10}=-n_2 \!\cdot\! (k_1-k_2)\,,
& &\cD_{11}=-n_2 \!\cdot\! (k_3-k_2)\,,
& &\cD_{12}=-n_2 \!\cdot\! k_3\,, & \\
&\quad     \cD_{13} =-n_J \!\cdot\!  k_1 - \omega\,,
& &\cD_{14}=-n_J \!\cdot\!  k_2 - \omega\,,
& &\cD_{15}=-n_J \!\cdot\!  k_3 - \omega\,.&
\nn\\ 
&\text{Family 6:}
\nn\\
&\quad    \cD_7=-n_1 \!\cdot\!  k_1\,, 
& &\cD_8=-n_1 \!\cdot\!  (k_1-k_2)\,, 
& & \cD_9= -n_1 \!\cdot\! (k_3-k_2)\,, & \nn\\
&\quad    \cD_{10}=-n_2 \!\cdot\! k_2\,,
& &\cD_{11}=-n_2 \!\cdot\!  k_3\,,
& &\cD_{12}=-n_2 \!\cdot\! (k_3-k_1)\,, & \\
&\quad     \cD_{13} =-n_J \!\cdot\!  k_1 - \omega\,,
& &\cD_{14}=-n_J \!\cdot\!  k_2 - \omega\,,
& &\cD_{15}=-n_J \!\cdot\!  k_3 - \omega\,.&
\nn\\ 
&\text{Family 7:}
\nn\\
&\quad    \cD_7=-n_1 \!\cdot\!  k_1\,, 
& &\cD_8=-n_1 \!\cdot\!  (k_1-k_3)\,, 
& & \cD_9= -n_1 \!\cdot\! (k_2-k_3)\,, & \nn\\
&\quad    \cD_{10}=-n_2 \!\cdot\! k_1\,,
& &\cD_{11}=-n_2 \!\cdot\!  (k_1-k_2)\,,
& &\cD_{12}=-n_2 \!\cdot\! (k_1-k_3)\,, & \\
&\quad     \cD_{13} =-n_J \!\cdot\!  k_1 - \omega\,,
& &\cD_{14}=-n_J \!\cdot\!  (k_1-k_3) - \omega\,,
& &\cD_{15}=-n_J \!\cdot\!  (k_2-k_3) - \omega\,.&
\nn\\ 
&\text{Family 8:}
\nn\\
&\quad    \cD_7=-n_1 \!\cdot\!  k_1\,, 
& &\cD_8=-n_1 \!\cdot\!  k_2\,, 
& & \cD_9= -n_1 \!\cdot\! k_3\,, & \nn\\
&\quad    \cD_{10}=-n_2 \!\cdot\! (k_2-k_1)\,,
& &\cD_{11}=-n_2 \!\cdot\!  (k_3-k_1)\,,
& &\cD_{12}=-n_2 \!\cdot\! k_3\,, & \\
&\quad     \cD_{13} =-n_J \!\cdot\!  (k_3-k_1) - \omega\,,
& &\cD_{14}=-n_J \!\cdot\!  k_2 - \omega\,,
& &\cD_{15}=-n_J \!\cdot\!  k_3 - \omega\,.&
\nn\\ 
&\text{Family 9:}
\nn\\
&\quad    \cD_7=-n_1 \!\cdot\!  k_1\,, 
& &\cD_8=-n_1 \!\cdot\!  (k_3-k_1)\,, 
& & \cD_9= -n_1 \!\cdot\! (k_3-k_2)\,, & \nn\\
&\quad    \cD_{10}=-n_2 \!\cdot\! k_1\,,
& &\cD_{11}=-n_2 \!\cdot\!  (k_3-k_1)\,,
& &\cD_{12}=-n_2 \!\cdot\! (k_2-k_1)\,, & \\
&\quad     \cD_{13} =-n_J \!\cdot\!  k_3 - \omega\,,
& &\cD_{14}=-n_J \!\cdot\!  (k_3-k_1) - \omega\,,
& &\cD_{15}=-n_J \!\cdot\!  (k_3-k_2) - \omega\,.&
\nn\\ 
&\text{Family 10:}
\nn\\
&\quad    \cD_7=-n_1 \!\cdot\!  (k_1-k_2)\,, 
& &\cD_8=-n_1 \!\cdot\!  (k_3-k_2)\,, 
& & \cD_9= -n_1 \!\cdot\! k_3\,, & \nn\\
&\quad    \cD_{10}=-n_2 \!\cdot\! k_1\,,
& &\cD_{11}=-n_2 \!\cdot\!  (k_1-k_2)\,,
& &\cD_{12}=-n_2 \!\cdot\! k_3\,, & \\
&\quad     \cD_{13} =-n_J \!\cdot\!  k_1 - \omega\,,
& &\cD_{14}=-n_J \!\cdot\!  (k_1-k_2) - \omega\,,
& &\cD_{15}=-n_J \!\cdot\!  k_3 - \omega\,.& 
\nn\\
&\text{Family 11:}
\nn\\
&\quad    \cD_7=-n_1 \!\cdot\!  (k_1-k_2)\,, 
& &\cD_8=-n_1 \!\cdot\!  k_2\,, 
& & \cD_9= -n_1 \!\cdot\! (k_1-k_3)\,, & \nn\\
&\quad    \cD_{10}=-n_2 \!\cdot\! (k_1-k_2)\,,
& &\cD_{11}=-n_2 \!\cdot\!  (k_3-k_2)\,,
& &\cD_{12}=-n_2 \!\cdot\! k_3\,, & \\
&\quad    \cD_{13} =-n_J \!\cdot\!  k_1 - \omega\,,
& &\cD_{14}=-n_J \!\cdot\!  k_2 - \omega\,,
& &\cD_{15}=-n_J \!\cdot\!  (k_1-k_2+k_3) - \omega\,.& \nn
\end{align} 

\section{Feynman parameter integrations with  HyperInt}
\label{app:HyperIntExample}
\begin{figure}
	\begin{center}
		\includegraphics[width=0.26\textwidth]{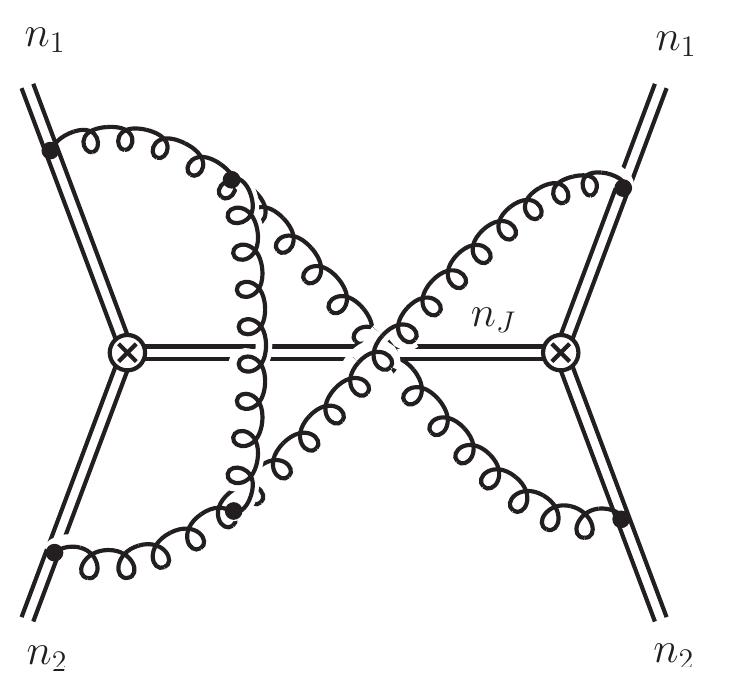}
	\end{center}
	\caption{\label{fig:feyndia8} Three-loop Feynman diagram with imaginary part for $n_{12}>0$.}
\end{figure}
Here we demonstrate the calculation of the three-loop MIs
with an imaginary part related to a physical threshold (similar to the two-loop example in \sec{disc}) using $\texttt{HyperInt}$~\cite{Panzer:2014caa}.
As a concrete example we consider one of the most complicated MIs of this kind, namely the integral $G^{[6]}_{1,1,0,1,1,1,1,0,1,1,0,1,0,0,1}$ in family~6,%
\footnote{Here we use the notation $G^{[6]}_{a_1,\ldots,e_3} \equiv G(a_1,\ldots,e_3)$ for the integrals in family 6 defined according to \eq{Gintegral}.}
which contributes to the Feynman diagram in fig.~\ref{fig:feyndia8}. 
The corresponding $\mathcal{U}$ and $\mathcal{F}$ polynomials in the Feynman parameter ($x_i$) representation, see e.g.~\rcite{Smirnov:2012gma}, are given by
\begin{align}
\mathcal{U}={}& x_1 x_2 x_5+x_1 x_4 x_5+x_2 x_4 x_5+x_1 x_6 x_5+x_2 x_6 x_5+x_1 x_2 x_6+x_1 x_4 x_6+x_2 x_4 x_6\,,\\
\mathcal{F}={}& \frac{n_{12}}{2}  x_4 x_5 x_7 x_{10}+\frac{n_{12}}{2}  x_4 x_6 x_7 x_{10}
+\frac{n_{12}}{2}  x_5 x_6 x_7 x_{10}-\frac{n_{12}}{2}  x_1 x_6 x_9x_{10}
-\frac{n_{12}}{2}  x_2 x_5 x_7 x_{12} \nn\\
&+\frac{n_{12}}{2}  x_1 x_2 x_9 x_{12}+\frac{n_{12}}{2}  x_1 x_4 x_9 x_{12}+\frac{n_{12}}{2}  x_2 x_4x_9 x_{12}+x_1 x_5 x_{10} x_{15}+x_4 x_5 x_{15} x_{10} \nn\\
&+x_5 x_6 x_{10} x_{15} +x_1 x_2 x_5 x_{15}+x_1 x_4 x_5 x_{15}+x_2 x_4
x_5 x_{15}+x_1 x_2 x_6 x_{15}+x_1 x_4 x_6 x_{15} \nn\\
&+x_2 x_4 x_6 x_{15}+x_1 x_5 x_6 x_{15}+x_2 x_5 x_6 x_{15}+x_4 x_5 x_7 x_{15}+x_2 x_6 x_7 x_{15}+x_4 x_6 x_7 x_{15} \nn\\
&+x_5 x_6 x_7 x_{15}+x_1 x_2 x_9 x_{15}+x_1 x_4 x_9 x_{15}+x_2 x_4 x_9 x_{15}+x_2 x_6 x_9 x_{15}+x_1 x_2 x_{12} x_{15} \nn\\
&+x_1 x_4 x_{12}x_{15}+x_2 x_4 x_{12} x_{15}+x_1 x_5 x_{12} x_{15}+x_4 x_6 x_{10} x_{15}  \,,
\label{eq:Fpoly}
\end{align}
where we set $n_1^2=n_2^2=n_J^2=0$, $n_{1J}=n_{2J}=2$, $\omega=-1$ for convenience.
The dependence on these variables can be easily restored according to \eq{Gres}.
The  index $i$ of the $x_i$ refers to the corresponding propagator in family 6.
Note that our definition of the integrals in \eq{Gintegral} implies that an infinitesimal $-\ri 0$ is to be added to $\mathcal{F}$ in \eq{Fpoly}. 

By a dimensional recurrence relation, $G^{[6]}_{1,1,0,1,1,1,1,0,1,1,0,1,0,0,1}$ can be related to the integral $G^{[6]}_{2,1,0,2,2,1,1,0,1,1,0,2,0,0,4}$, which is finite in $d=8-2\epsilon$ dimensions, and less complicated already known integrals.
The coefficients of the integrals in such relations are real and only depend on $d$.
Regardless of the sign of $n_{12}$, the $\mathcal{F}$ polynomial becomes negative in certain integration regions.
We now exploit the projective nature of the Feynman parameter representation~\cite{Smirnov:2012gma} by fixing $x_4=1$ 
and are left with the integrations over the remaining $x_i$ from $0$ to $+\infty$. 
For $n_{12}=2$ and dropping the infinitesimal $-\ri0$ $\texttt{HyperInt}$ manages to perform the 
$x_7,x_9,x_{15},x_{10},x_{12},x_1,x_2,x_5,x_6$ integrations in that order. Because we have removed the infinitesimal $-\ri0$,
$\texttt{HyperInt}$ automatically adds an imaginary infinitesimal $\delta_{x_6} \ri \varepsilon$ with $\delta_{x_6}=\pm 1$ to the parameter $x_6$ in order to deform the integration contour away from an (integrable) threshold singularity for $\mathcal{F} \to 0$.
Note that, in contrast to the original integral with fixed $-\ri 0$,
 the sign of the imaginary infinitesimal now depends on the integration region.
The $\texttt{HyperInt}$ output will therefore in general differ from the integral we aim to compute.
This difference however only affects the imaginary part of the integral, which is irrelevant for our purpose as discussed in \sec{calc}.
The real part agrees with the original one.
To see this we write
\begin{align}
  \mathrm{Re}\bigl[ G(\vec{a}, \vec{b},\vec{c}, \vec{e},\epsilon) \bigr]
  ={}& \frac{1}{2} \Bigl( G(\vec{a}, \vec{b},\vec{c}, \vec{e},\epsilon) \big|_{+\ri0} 
  + G(\vec{a}, \vec{b},\vec{c}, \vec{e},\epsilon) \big|_{-\ri0} \Bigr)\nn\\
  ={}&  \frac{1}{2} \Bigl( G(\vec{a}, \vec{b},\vec{c}, \vec{e},\epsilon) \big|_{+\ri0}^>
  + G(\vec{a}, \vec{b},\vec{c}, \vec{e},\epsilon) \big|_{+\ri0}^< \nn\\
  &+ G(\vec{a}, \vec{b},\vec{c}, \vec{e},\epsilon) \big|_{-\ri0}^>
  + G(\vec{a}, \vec{b},\vec{c}, \vec{e},\epsilon) \big|_{-\ri0}^< 
  \Bigr) \nn\\
  ={}&  \frac{1}{2} \Bigl( G(\vec{a}, \vec{b},\vec{c}, \vec{e},\epsilon) \big|_{+\delta_{x_k} \ri \varepsilon}^>
  + G(\vec{a}, \vec{b},\vec{c}, \vec{e},\epsilon) \big|_{-\delta_{x_k}\! \ri \varepsilon}^< \nn\\
  &+ G(\vec{a}, \vec{b},\vec{c}, \vec{e},\epsilon) \big|_{-\delta_{x_k}\! \ri \varepsilon}^>
  + G(\vec{a}, \vec{b},\vec{c}, \vec{e},\epsilon) \big|_{+\delta_{x_k}\! \ri \varepsilon}^<
  \Bigr) \nn\\
  ={}& \mathrm{Re} \Bigl[ G(\vec{a}, \vec{b},\vec{c}, \vec{e},\epsilon) \big|_{+\delta_{x_k}\! \ri \varepsilon}^>
  + G(\vec{a}, \vec{b},\vec{c}, \vec{e},\epsilon) \big|_{+\delta_{x_k}\! \ri \varepsilon}^< 
  \Bigr] \nn\\
  ={}& \mathrm{Re} \Bigl[ G(\vec{a}, \vec{b},\vec{c}, \vec{e},\epsilon) \big|_{+\delta_{x_k}\! \ri \varepsilon} \Bigr]
  \label{eq:Reargument}
\end{align}
for a generic integral $G$.%
\footnote{On the other hand we have
$\mathrm{Im}\Bigl[ G(\vec{a}, \vec{b},\vec{c}, \vec{e},\epsilon)\big|_{-\ri0} \Bigr] = 
\mathrm{Im} \Bigl[ G(\vec{a}, \vec{b},\vec{c}, \vec{e},\epsilon) \big|_{-\delta_{x_k}\!\ri \varepsilon}^> \Bigr]
- \mathrm{Im} \Bigl[
  + G(\vec{a}, \vec{b},\vec{c}, \vec{e},\epsilon) \big|_{-\delta_{x_k}\!\ri \varepsilon}^< \Bigr]$.}
The subscript indicates whether the integral is evaluated with fixed $\pm \ri 0$ or an imaginary infinitesimal $\delta_{x_k} \ri \varepsilon$ added to the Feynman parameter $x_k$ in the $\mathcal{F}$ polynomial. 
The $>$ ($<$) superscript indicates the contribution from all integration regions, where the coefficient of $x_k$ (and therefore of $\delta_{x_k} \ri \varepsilon$) is positive (negative).
We have checked \eq{Reargument} by computing our quasi-finite integrals for different $x_i$ initially fixed to one and different integration orders, which causes $\texttt{HyperInt}$ to assign the $\delta_{x_k} \ri \varepsilon$ to different $x_k$.
In most cases we found a configuration, where $\texttt{HyperInt}$ adds the imaginary infinitesimal to a Feynman parameter with positive definite coefficient, which is equivalent to a fixed $-\ri 0$ in the $\mathcal{F}$ polynomial.
In this way we explicitly confirmed that it is irrelevant for the real part of the result whether the coefficient of the $\delta_{x_k} \ri \varepsilon$ is positive or varies in sign depending on the integration region.

With the output from $\texttt{HyperInt}$ we obtain for our concrete example in $d=8-2\epsilon$ dimensions ($\delta _{x_6}^2=1$),
\begin{align}
\mathrm{Re}\Bigl[ G^{[6]}&_{2,1,0,2,2,1,1,0,1,1,0,2,0,0,4} \Bigr] =
\left(\frac{1}{72}+\frac{7 \pi ^2}{72}\right) \zeta _3-\frac{25 \zeta _5}{72}-\frac{29 \pi^4}{8640}-\frac{11 \pi ^2}{432}+\frac{\pi ^2 }{48} \delta _{x_6}^2
\nn\\
&+\Bigg[\frac{\pi ^2}{18}  \zeta _{1,-3}+\frac{17 \zeta _3^2}{288}-\frac{611 \pi ^2 \zeta _3}{1728}+\frac{221 \zeta _3}{216}-\frac{1693 \zeta_5}{1728}
+\frac{2021\pi ^6}{544320}-\frac{\pi ^4}{64}-\frac{823 \pi ^2}{5184}
\nn\\
&\qquad +\frac{7\pi ^2}{24}  \ln 2+\left(\frac{3 \pi ^2 \zeta _3}{16}-\frac{\pi ^6}{540}+\frac{\pi ^4}{288}+\frac{17 \pi ^2}{192}\right) \delta _{x_6}^2\Bigg]\eps
\nn\\
&+\Bigg[\frac{7 \zeta _3 }{36}\zeta _{1,-3}-\frac{257\pi ^2 }{432} \zeta _{1,-3}-\frac{7 }{12}\zeta _{1,-3}+\frac{28}{51} \zeta _{1,1,-5}-\frac{ \pi ^2}{18}\zeta _{1,1,-3}-\frac{14}{153} \zeta _{1,3,-3}
\nn\\
&\qquad
-\frac{10417 \zeta _3^2}{6912}+\frac{29215 \pi ^4 \zeta _3}{88128}-\frac{1937 \pi ^2 \zeta_3}{2592}+\frac{5627 \zeta _3}{1296}+\frac{30181 \pi ^2 \zeta _5}{58752}
\nn\\
&\qquad
-\frac{2111 \zeta _5}{1296} -\frac{335531 \zeta _7}{39168}
-\frac{53731 \pi ^6}{1306368}+\frac{3869 \pi ^4}{25920}
-\frac{14965 \pi^2}{31104}
\nn\\
&\qquad 
+\left(\frac{49\zeta _3}{48}+\frac{71 \pi ^2}{48}\right) \ln 2+\frac{7\pi ^2}{48}  \ln^2 2+\left(-\frac{\pi ^4}{4}  \zeta _3+\frac{\pi ^6}{48}-\frac{\pi ^4}{24}\right) \delta_{x_6}^4
\nn\\
&\qquad+\left(\frac{29 \pi ^4 \zeta _3}{96}+\frac{9 \pi ^2 \zeta _3}{16}+\frac{7 \pi ^2 \zeta _5}{32}-\frac{113 \pi ^6}{103680}-\frac{187\pi ^4}{864}+\frac{355 \pi ^2}{1152}\right) \delta _{x_6}^2
\Bigg]\eps^2
\nn\\
&+\cO(\eps^3)\,,
\end{align}
where we give all terms in the $\eps$ expansion required for our soft function calculation.
Note that the multiple zeta values, the $\ln 2$ terms, and the terms of transcendental weight $>6$ cancel among the different MIs and do not appear in the final result for the soft function.

\phantomsection
\addcontentsline{toc}{section}{References}
\bibliographystyle{jhep}
\bibliography{VJthrsoft}

\end{document}